\documentclass[11pt,a4paper]{amsart}

\usepackage[latin1]{inputenc}
\usepackage[english]{babel}
\usepackage{amsmath}

\usepackage{amssymb, mathabx}
\usepackage[numbers]{natbib}
\usepackage{graphicx}
\usepackage{braket}
\usepackage[pdftex,plainpages=false,colorlinks,hyperindex,bookmarksopen,linkcolor=red,citecolor=blue,urlcolor=blue]{hyperref}

\usepackage{mathrsfs}

\usepackage{epstopdf}

\usepackage{braket}

\bibpunct{[}{]}{;}{n}{,}{,}
\usepackage[hmargin=3cm,vmargin={3.5cm,4cm}]{geometry}

\theoremstyle{theorem}
\newtheorem{thm}{Theorem}

\theoremstyle{definition}                                 %stile corsivo
\theoremstyle{definition}                           %stile roman
\theoremstyle{remark}                             %stile per osservazioni
\newtheorem*{rmk}{Remark}              %definizione ambiente osservazione

\usepackage{color}

\newcommand{\be}{\begin{eqnarray}}
\newcommand{\ee}{\end{eqnarray}}

\newcommand{\R}{\mathbb{R}}  %%%%% \R = \mathbb{R}.
\newcommand{\N}{\mathbb{N}} %%%% \K = \mathbb{K}.
\def\eg{{\it e.g. }} 
\def\ie{{\it i.e. }}

\newcommand{\wt}[1]{\widetilde{#1}}

\newcommand{\opO}{\mathscr{O}}
\newcommand{\opL}{\mathbb{L}}

\def\d{\partial}

\def\ds{\displaystyle}
\def\ds{\displaystyle}
\def\eg{{\it e.g.}\ }
\def\ie{{\it i.e.}\ }

\numberwithin{equation}{section}

\allowdisplaybreaks

\begin{document}
\title{On Transient Waves in Linear Viscoelasticity}

    \author{Ivano Colombaro$^1$}
		\address{${}^1$ Department of Information and Communication Technologies, Universitat Pompeu Fabra and INFN.
		C/Roc Boronat 138, Barcelona, SPAIN.}
		\email{ivano.colombaro@upf.edu}
	
	    \author{Andrea Giusti$^2$}
		\address{${}^2$ Department of Physics $\&$ Astronomy, University of 	
    	    Bologna and INFN. Via Irnerio 46, Bologna, ITALY and 
	    	 Arnold Sommerfeld Center, Ludwig-Maximilians-Universit\"at, 
	    	 Theresienstra{\ss}e~37, 80333 M\"unchen, GERMANY.   
    	    }
		\email{andrea.giusti@bo.infn.it}
	
    \author{Francesco Mainardi$^3$}
    	    \address{${}^3$ Department of Physics $\&$ Astronomy, University of 	
    	    Bologna and INFN. Via Irnerio 46, Bologna, ITALY.}
			\email{francesco.mainardi@bo.infn.it}


    \thanks{Paper published in \textbf{Wave Motion 74C (2017) pp. 191--212}, \textbf{DOI}: \href{http://www.sciencedirect.com/science/article/pii/S0165212517300963}{10.1016/j.wavemoti.2017.07.008}.}

    \date  {\today}%%{January 2016}

\begin{abstract}
	The aim of this paper is to present a comprehensive review of method of the wave-front expansion, also known in the literature as the Buchen-Mainardi algorithm. In particular, many applications of this technique to the fundamental models of both ordinary and fractional linear viscoelasticity are thoroughly presented and discussed.
\end{abstract}

    \maketitle

    \tableofcontents

\newpage

\section{Introduction} \label{Sec-1}
	In the last three decades, Linear Viscoelasticity \cite{Coleman, extra, Gurtin-Sternberg, Hanyga, Mainardi_BOOK10} has presented itself as a fundamental tool for modelling various physical situations, from seismology to biophysics (see \eg \cite{BM-ZAMP, AG-FM_MECC16, MST}). In this framework, models involving fractional derivatives in the constitutive equations play an important role, allowing for a natural formal background for describing systems with naturally occurring memory effects (see \eg \cite{Fabrizio-Morro, Gurtin-Sternberg, Mainardi_BOOK10, Mainardi-Spada 2011, Marchetta, Marchetta-2, Marchetta-3}). There are also a few models that show a continuous transition from a purely fractional nature to an ordinary one, which appear to be of particular interest for their applications in hemodynamics (see \cite{IC-AG-FM-2016, AG-FM_MECC16}). 
	
	An important phenomenon that tend to emerge in various branches of physics is the appearance of transient effects, \ie processes that lead to a burst of energy in a system, caused by a sudden change of state, that last for a very short time. These processes emerge in the description of mechanical system as transient waves. It is therefore very important to understand how these short-lived waves propagate within certain media. 
	
	The first attempt to formally describe the behaviour of the wave-front of a transient wave, as it propagates in a viscoelastic material, was proposed in the seventies by Buchen and Mainardi, with the seminal paper \cite{Buchen-Mainardi 1975}. Recently, the so called formalism of the \textit{Buchen-Mainardi wave-front expansion}, was applied to the Bessel models of linear viscoelasticity (see \cite{IC-AG-FM-2016}). It is worth remarking that the result of this analysis (see \cite{BM-ZAMP}) has shown a consistency between the model proposed in \cite{AG-FM_MECC16} by Giusti and Mainardi, and the expected behaviour of a pressure wave as it propagates within a large artery.
	
	It is worth remarking that, in recent years, many authors have tackled the problem of the wave-front expansion in viscoelastic media. Among the many who have worked on this matter (see \eg \cite{K, K1, MT, Pipkin}), we feel appropriate to highlight the remarkable results obtained by Hanyga and Seredynska, for further details see \eg \cite{Hanyga, Stronzo1, Stronzo3}.
	
	The aim of this paper is then to review and generalize the Buchen-Mainardi algorithm for the wave-front expansion in viscoelastic media. Specifically, after a thorough review of the algorithm, we show an explicit implementation of the formalism for all the main fundamental model of both ordinary and fractional linear viscoelasticity. The article is therefore organized as follows:
	
	In Section \ref{Sec-2}, we thoroughly revisit and generalize the Buchen-Mainardi algorithm.
	
	In Section \ref{Sec-3} and \ref{Sec-4} we discuss the formalism for the Maxwell model and to the fractional Maxwell model of general order $\alpha$. Then, we explicitly solve the fractional case for two specific realizations of the model, precisely $\alpha = 1/2 , \, 3/4$.

	In Section \ref{Sec-5} and \ref{Sec-6} we then lay the groundwork for the application of the formalism to the Voigt model and to the fractional Voigt model of general order $\alpha$. Then, for sake of brevity, we explicitly present the discussion for the most interesting fractional realization of the model, \ie $\alpha = 1/2$.
	
	Finally, we conclude the paper presenting some final considerations.
	
%%%%BUCHEN-MAINARDI
%%%%%%%

\section{Buchen-Mainardi wave-front expansion} \label{Sec-2}
	In this section we wish to clarify and slightly generalize the renown algorithm for performing asymptotic wave-front expansions for singular viscoelastic models, first presented by Buchen and Mainardi in \cite{Buchen-Mainardi 1975}.
	Let us consider an homogeneous and isotropic viscoelastic medium of density $\rho$. This body is assumed to be semi-infinite in extent (\ie $x\geq 0$) and unperturbed for $t < 0$.
	
	For $t \geq 0$, the accessible portion of the body (\ie $x= 0$) is subjected to a perturbation (an input signal) denoted by $r_0 (t)$. The problem now is to compute the response $r(t,x)$ of the material for $x > 0$. This response function can be either the stress or the strain, or some other quantities that depend on the specific implementation of the model (for example, in \cite{Buchen-Mainardi 1975} the response function could also represent the displacement $u(t,x)$ or the particle velocity $\dot u (t, x)$). 
	
	Let $J(t)$ be the uniaxial creep compliance corresponding to our viscoelastic body, for which we set $J_0 \equiv J(0^+)$. Moreover, defining the wave-front velocity as
	$$ c = \frac{1}{\sqrt{\rho \, J_0}} \, , $$
and denoting with $\Psi (t)$ the creep memory function, defined as
	$$ \Psi (t) = \frac{1}{J_0} \frac{dJ(t)}{dt} \, , \qquad \mbox{as} \,\,\, t > 0 \, , $$
thus we have that the equation of motion reads (see \eg \cite{Mainardi_BOOK10})
\begin{equation} \label{eq-motion}
\frac{\partial ^2 r}{\partial x^2} - \frac{1}{c^2} \, \Big[ 1 + \Psi (t) \ast \, \Big] \frac{\partial ^2 r}{\partial t^2} = 0
\end{equation}  
where the $\ast$ denotes the convolution in the Laplace sense.

	Now, taking the Laplace transform of Eq.~(\ref{eq-motion}), switching from the time domain to the Laplace domain, we get the following second order differential equation,
	\begin{equation} \label{eq-motion-laplace}
	\left[ \frac{\partial ^2}{\partial x^2} - \mu ^{2} (s) \right] \wt{r} (s, x) = 0 \, ,
	\end{equation}
where	
	\begin{equation}
	\mu (s) \equiv s \, \left[ \rho \, s \, \wt{J} (s) \right] ^{1/2} \, .
	\end{equation}
Taking into account the boundary conditions, one can easily find a formal solution for Eq.~(\ref{eq-motion-laplace}) in the Laplace domain, indeed
	\begin{equation} \label{eq-sol-laplace}
	\wt{r} (s, x) = \wt{r} _{0} (s) \, \exp \left[ - x \, \mu (s) \right] \, .
	\end{equation} 
	
	As discussed in \cite{Buchen-Mainardi 1975}, the aim of this procedure is to compute the asymptotic expansion of $r(t, x)$ in the neighborhood of the pulse onset. This is done by inverting, term by term, the expansion of (\ref{eq-sol-laplace}) as $s \to \infty$ that, taking profit of Watson's lemma or by means of the Tauberian theorems (see \eg \cite{Miller}), will correspond to an expansion for $t \to (x/c)^{+}$.

	Let us first discuss the behavior of $\mu (s)$ as $s \to \infty$. For a singular model it is known (see \eg \cite{Mainardi_BOOK10}) that $\mu (s)$ will have an expansion in terms of decreasing powers of $s$. Thus, in a general fashion, one could write this expansion as   
	\begin{equation} \label{eq-mu-as}
	\mu (s) \overset{s \to \infty}{\sim} \sum _{k=0} ^{\infty} b_{k} \, s^{1 - \beta _{k}} \, , \qquad 0 = \beta _{0} < \beta _{1} < \cdots
	\end{equation} 
	Now, it is important to define two quantities whose relevance will appear clear later in this discussion. Then, let us denote $\mu _{+} (s)$ the sum of the first $m+1$ ($m \in \N_{0}$) terms of Eq.~(\ref{eq-mu-as}) that depend on positive powers of $s$, \ie
	\begin{equation} \label{eq-mu-plus}
	\mu _{+} (s) \equiv \sum _{k=0} ^{m} b_{k} \, s^{1 - \beta _{k}} \, , \qquad \beta _{k} \leq 1 \, , \,\, k = 0, 1, \ldots, m \, .
	\end{equation}
 	Moreover, we denote with $\mu _{-} (s)$ the remainder of the series (\ref{eq-mu-as}), \ie
	\begin{equation} \label{eq-mu-minus}
	\mu _{-} (s) \equiv \sum _{k=0} ^{\infty} b_{k} \, s^{1 - \beta _{k}} - \mu _{+} (s) \, .
	\end{equation}
	Therefore, taking profit of these definitions, we can rewrite the solution in Eq.~(\ref{eq-sol-laplace}), for $s \to \infty$, as follows
	\begin{equation} \label{eq-sol-laplace-as}
	\wt{r} (s, x) 
	\overset{s \to \infty}{\sim} 
	\wt{r} _{0} (s) \, \exp \left[ - x \, \mu _{+} (s) \right] \, \wt{R} (s, x) \, , 
	\end{equation}
with
\begin{equation}
\wt{R} (s, x) \equiv \exp \left[ - x \, \mu _{-} (s) \right] \, .
\end{equation}
	According to \cite{Buchen-Mainardi 1975, Mainardi_BOOK10}, we shall refer to $\mu _{+} (s)$ as the \textit{principal part} of the expansion of $\mu (s)$, which is directly connected to the asymptotic expansion of the creep compliance, either in the Laplace domain or in the time domain. Taking profit of the general asymptotic expansion of the creep compliance for a singular viscoelastic model, \ie
	$$ J(t) \sim J_{0} + O (t^{\alpha}) \, , \qquad \mbox{as} \,\, t \to 0 ^{+} \,\, (\Leftrightarrow \,\, s \to \infty) \, , $$
 with $0 < \alpha \leq 1$, one can easily show that
	\begin{equation}
	\left\{
	\begin{aligned}
	& J_{0} = 0 \quad \Rightarrow  \quad b_{0} = 0 \, , \,\, \beta _{1} = \alpha / 2 \, ;\\
	& J_{0} > 0 \quad \Rightarrow  \quad b_{0} = 1/ c \, , \,\, \beta _{1} = \alpha \, .
	\end{aligned}
	\right.
	\end{equation}
	
	It is worth remarking that this restriction on the values of $\alpha$ allows for a finite propagation speed of the wave-front, as discussed in \cite{Stronzo1}.
	
	Given that
	\begin{equation}
	\mu _{-} (s) = \sum _{k=m+1} ^{\infty} b_{k} \, s^{1 - \beta _{k}} \, , \qquad \mbox{with} \,\, \beta _{k} > 1 \, , \,\, k = m+1, \, m+2, \ldots
	\end{equation} 
	we can expand $\wt{R} (s, x)$ in terms of decreasing powers of $s$ as follows,
	\begin{equation} \label{eq-expansion-R}
	\wt{R} (s, x) \overset{s \to \infty}{\sim} \sum _{k=0} ^{\infty} v _{k} (x) \, s^{- \lambda _{k}} \, , \qquad 0 \leq \lambda _{0} < \lambda _{1} < \cdots \, ,
	\end{equation}
	for some functions $v_{k} (x)$ that will encode the dependence on the spatial direction.
	
	Then, plugging the expression in Eq.~(\ref{eq-sol-laplace-as}) into Eq.~(\ref{eq-motion-laplace}) one finds that $\wt{R} (s, x)$ should also satisfy the following differential equation
	\begin{equation} \label{eq-L}
	\opO \, \wt{R} (s, x) = \left\{ \frac{d^{2}}{dx^{2}} - 2 \, \mu _{+} (s) \, \frac{d}{dx} - [ \mu ^2 (s) - \mu _+ ^2 (s) ]  \right\} \, \wt{R} (s, x) = 0 \, , 
	\end{equation}
	together with the condition
	\begin{equation}
	\wt{R} (s, 0) = 1 \, .
	\end{equation}
	
	In order to perform such an expansion for $\wt{R} (s, x)$ we should take profit of the following theorem by Friedlander and Keller, that states
	\begin{thm}[\cite{FK}]
	Let $\opL$ be a linear differential operator such that its asymptotic expansion with respect to a parameter $\varepsilon$, as $\varepsilon \to 0$, is given by
	$$ \opL \overset{\varepsilon \to 0}{\sim} \sum _{i = 0} ^{\infty} \varepsilon ^{\nu _{i}} \, \opL _{i} \, , \qquad 0 = \nu _{0} < \nu _{1} < \cdots \, , $$
	where $\opL _{i}$ are some suitable linear differential operators.
	
	Let also assume $f$ to be a solution of
	$$ \opL f = 0 \, , $$
	with an asymptotic expansion as $\varepsilon \to 0$
	$$ f \overset{\varepsilon \to 0}{\sim} \sum _{k=0} ^{\infty} \varepsilon ^{\lambda _{k}} \, v_{k} \, , \qquad \lambda _{0} < \lambda _{1} < \cdots \, .$$
	
	If the asymptotic expansion of $\opL$ is given by termwise application of the $\opL _{i}$ to the $v_{j}$ and if $\opL _{0} v_{k} \neq 0$ for $k > 0$, then the coefficients $v_{k}$ satisfy the following recursive system of equations
	\begin{equation*}
	\begin{aligned}
	& \opL _{0} \, v_{0} = 0 \, , \\
	& \opL _{0} \, v_{k} = - \sum _{i j} \, \opL _{i} \, v_{j} \, , \qquad k = 1, 2, 3 \ldots \, ,
	\end{aligned}
	\end{equation*}
	where the summation is taken over values of $i, \, j$ such that $\nu _{i} + \lambda _{j} = \lambda _{k}$. The value of $\lambda _{0}$ is arbitrary, but $\lambda _{k}$ for $k>0$ is the $(k+1)$-st number in the increasing sequence formed from the set $\lambda _{0} + \sum _{i=1} ^{\infty} m_{i} \, \nu _{i}$, where $m_{i}$ are non-negative integers.
	\end{thm}
	
	\begin{rmk}
	Notice that $j$ is a dummy index, indeed from the condition $\nu _{i} + \lambda _{j} = \lambda _{k}$ one can easily infer that $j = j(i, k)$. Therefore, one can rewrite the previous system as
	\begin{equation*}
	\begin{aligned}
	& \opL _{0} \, v_{0} = 0 \, , \\
	& \opL _{0} \, v_{k} = - \sum _{i} \, \opL _{i} \, v_{j (i, k)} \, , \qquad k = 1, 2, 3 \ldots \, ,
	\end{aligned}
	\end{equation*}
	for some function $j$ of $i$ and $k$ defined by the relation between the exponents of the asymptotic expansions.
	\end{rmk}
	
	If we now expand $\mu (s)$ and $\mu _{+} (s)$ in powers of $s$, as $s \to \infty$, in Eq.~(\ref{eq-L}) and then divide the whole equation for the term containing the highest power of $s$, we get a differential operator of the form required by the previous theorem. In particular, for these specific application we will have that $\varepsilon = 1/s$ (which actually goes to zero as $s \to \infty$) and the operators $\opL _{i}$ will be some functions of the derivative with respect to $x$. The functional dependence on $\partial /\partial x$ of these operators will depend on the expansion of $\mu (s)$.
	
	If we implement the previous discussion, we obtain quite generally
	\begin{equation*}
	\begin{aligned}
	& \opL _{0} = \frac{\partial}{\partial x} \, , \\
	& \opL _{i} = p_{i} \, \frac{\partial ^{2}}{\partial x ^{2}} + q_{i} \, \frac{\partial}{\partial x} + r_{i} \, , \qquad i = 1, 2, 3 \ldots \, ,
	\end{aligned}
	\end{equation*}
	where the coefficients $p _{i}$, $q_{i}$ and $r_{i}$, together with the exponents $\nu _{i}$, are determined by the asymptotic expansion of $\mu (s)$ and, therefore, from the asymptotic behavior of $s \, \wt{J} (s) $ as $s \to \infty$.

	Now, if we impose the condition $\wt{R} (s, 0) = 0$ on the asymptotic expansion of $\wt{R} (s, x)$ in Eq.~(\ref{eq-expansion-R}) we infer that
	\begin{equation}
	\lambda _{0} = 0 \, , \qquad v_{k} (0) = \delta _{k 0} \, ,
	\end{equation}
where $\delta _{k h}$ is the usual Kronecker delta function.

	Hence, from the theorem above we get that 
	\begin{equation}
	\lambda _{k} = \sum _{i = 1} ^{N} m_{i} \, \nu _{i} \, ,
	\end{equation}
	where $N$ is the label of the last non-vanishing sub-operators $\opL _{i}$ in the asymptotic expansion of $\opL$, and that the coefficients $v_{k} (x)$ will be given by the recursive system of equations
	\begin{equation} \label{eq-system-v}
	\frac{\partial v_{0}}{\partial x} = 0 \, , \qquad
	\frac{\partial v_{k}}{\partial x} = - \sum _{i} \left( p_{i} \, \frac{\partial ^{2}}{\partial x ^{2}} + q_{i} \, \frac{\partial}{\partial x} + r_{i} \right) v_{j (i, k)} \, , \qquad k = 1, 2, 3 \ldots \, ,
	\end{equation} 
where the function $j(i, k)$ is deduced by the condition $\nu _{i} + \lambda _{j} = \lambda _{k}$.

	It is easily understood that the solution for this system of equations is given by functions $v_k (x)$ such that $v_k (x) \in \R _k [x]$, \ie
	\begin{equation} \label{eq-poly}
	v_k (x) = \sum _{\ell = 0} ^k A_{k, \ell} \, \frac{x^\ell}{\ell !} 
	\, .
	\end{equation}
Plugging these polynomials into Eq.~(\ref{eq-system-v}) we get a recursive system for the coefficients $A_{k, \ell}$,
\begin{equation} \label{eq-system-A}
\begin{cases}
  A_{k, \ell} = \delta_{k \ell} \qquad \ell = 0 \, , \\
  A_{k, \ell} = - \ds \sum _{i} \left[ \sum _{J} \left( p_i A_{J, \, \ell+1} + q_i A_{J, \, \ell} + r_i A_{J, \, \ell-1} \right)\delta_{J, \, j(i,k)} \right] \qquad 1 \leq \ell \leq k \, , \\
  A_{k, \ell} = 0 \qquad \ell>k \, ,
 \end{cases}
\end{equation}	
where we decide to put a ``redundant'' sum over a dummy index $J$ in order to lighten the notation, avoiding cumbersome subscripts like $A_{j(i,k), \, \ell}$.

	Now, plugging the asymptotic expansion in Eq.~(\ref{eq-expansion-R}) into Eq.~(\ref{eq-sol-laplace-as}), we have that
	\begin{eqnarray}
	\wt{r} (s, x) 
	&\!\!\overset{s \to \infty}{\sim}\!\!& 
	\wt{r} _{0} (s) \, \exp \left[ - x \, \mu _{+} (s) \right] \, \wt{R} (s, x) =\\
	&\!\!=\!\!&
	\sum _{k=0} ^\infty v_k (x) \,
	\, \wt{r} _{0} (s) \, s^{- \lambda _k} \, 
	\exp \left[ - x \, \mu _{+} (s) \right] \, .
	\end{eqnarray}		  
	Then, defining
	\begin{equation} \label{eq-phi-laplace}
	\wt{\Phi} _k (s, x) \equiv \wt{r} _{0} (s) \, s^{- \lambda _k} \, 
	\exp \left[ - x \left( \mu _{+} (s) - \frac{s}{c} \right) \right] \, ,
	\end{equation}
with the convention $1/c = 0$ when $J_0 = 0$, we get
	\begin{equation} \label{eq-fin-laplace}
	\wt{r} (s, x) \overset{s \to \infty}{\sim}
	\exp \left[ - \frac{x \, s}{c} \right] \, 
	\sum _{k=0} ^\infty v_k (x) \, \wt{\Phi} _k (s, x) \, .
	\end{equation}
	For a mathematical discussion of the functions $\wt{\Phi} _k (s, x)$ we invite the interested reader to refer to \cite{Buchen-Mainardi 1975, Mainardi_BOOK10}. 
	
	Now, inverting the result in Eq.~(\ref{eq-fin-laplace}) back to the time domain (see \cite{Buchen-Mainardi 1975, Mainardi_BOOK10}) we get the asymptotic expansion for $r(t,x)$ as $t \to (x/c)^+$, \ie
	\begin{equation} \label{eq-fin}
	r (t, x) \,\, \overset{t \to (x/c)^+}{\sim} \,\, 
	\sum _{k=0} ^\infty v_k (x) \, \Phi _k \left(t - \frac{x}{c}, x \right) \, ,
	\end{equation}
where the functions $v_k (x)$ are computed by means of Eq.~(\ref{eq-poly}) and (\ref{eq-system-A}) and the functions $\Phi _k (t, x)$ are the Laplace inverse of the functions $\wt{\Phi} _k (s, x)$ in Eq.~(\ref{eq-phi-laplace}).

%%%%%%%	ORDINARY MAXWELL %%%%%%%
%%%%%%%%%%%%%%%%%%%%%%%%%%%

\section{The (Ordinary) Maxwell Model} \label{Sec-3}
	The ordinary Maxwell model of linear viscoelasticity is defined in terms of its constitutive equation
	\be
	\sigma (t) + a_{1} \, \frac{d \sigma (t)}{dt} = b_{1} \, \frac{d \varepsilon (t)}{dt} \, ,
	\ee
where $a_{1}$ and $b_{1}$ are strictly positive constants. 
	
	Applying the Laplace transform to both sides we get
	\be
	(1 + a_{1} \, s) \, \wt{\sigma} (s) = b_{1} \, s \, \wt{\varepsilon} (s) \, ,
	\ee
thus,
	\be
	\wt{\varepsilon} (s) = \frac{1 + a_{1} \, s}{b_{1} \, s} \, \wt{\sigma} (s) \, ,
	\ee
from which we infer that
\be
s \, \wt{J} _{M} (s) = \frac{1 + a_{1} \, s}{b_{1} \, s} \, .
\ee	
	It is interesting to remark that the creep compliance for the Maxwell model in the time domain is easily obtained by inverting the previous expression, indeed
	\be
	\wt{J} _{M} (s) = \frac{a_{1} \, s^{-1} + s^{-2}}{b_{1}} \,\, \div \,\, J_{M} (t) = J_{0} + J_{1} \, t \, ,
	\ee	
	where $J_{0} \equiv a_{1} / b_{1}$ and $J_{1} \equiv 1/b_{1}$.
	
	\subsection{Wave-front Expansion}
	For the (ordinary) Maxwell model we have that
	\be \label{eq-mu-maxwell}
	\mu (s) = \sqrt{\rho} \, s \, \left[s \, \wt{J} _{M} (s) \right] ^{1/2} = \sqrt{\rho} \, s \, \left(J_{0} + J_{1} \, s^{-1} \right) ^{1/2} = \frac{s}{c}\, \left( 1 + \frac{J_{1}}{J_{0}} \, s^{-1} \right) ^{1/2} \, ,
	\ee
	where we have taken profit of the relation $c = 1/ \sqrt{\rho \, J_{0}}$.
	
	Now, as $s \to \infty$ we have that
	\be
	\mu (s) &\!\!=\!\!& \frac{s}{c} \, \left( 1 + \frac{J_{1}}{J_{0}} \, s^{-1} \right) ^{1/2} = \frac{s}{c} \, \left[ 1 + \frac{J_{1}}{2 \, J_{0}} \, s^{-1} + \sum _{n = 2} ^{\infty} \binom {1/2} {n} \left( \frac{J_{1}}{J_{0}} \right)^{n} \, s^{-n} \right] =\\
	&\!\!=\!\!&
	\frac{s}{c} + \frac{J_{1}}{2 \, J_{0} \, c} + \frac{1}{c} \sum _{n = 2} ^{\infty} \binom {1/2} {n} \left( \frac{J_{1}}{J_{0}} \right)^{n} \, s^{1 - n} \, . \notag
	\ee
	From this expression one can easily read off $\mu _{+} (s)$ and $\mu _{-} (s)$, \ie
	\be
	\mu _{+} (s) &\!\!=\!\!& \frac{s}{c} + \frac{J_{1}}{2 \, J_{0} \, c} \, , \\
	\mu _{-} (s) &\!\!=\!\!& \frac{1}{c} \sum _{n = 2} ^{\infty} \binom {1/2} {n} \left( \frac{J_{1}}{J_{0}} \right)^{n} \, s^{1 - n} \, .
	\ee
	Then,
	\be
	\mu ^{2} (s) &\!\!=\!\!& \frac{s^{2}}{c^{2}} + \frac{J_{1}}{J_{0} \, c^{2}} \, s \, , \\
	\mu _{+} ^{2} (s) &\!\!=\!\!& \frac{s^{2}}{c^{2}} + \frac{J_{1}}{J_{0} \, c^{2}} \, s + \frac{J_{1} ^{2}}{4 \, J_{0} ^{2} \, c^{2}} \, ,
	\ee
	from which we get
	\be
	\mu ^{2} (s) - \mu _{+} ^{2} (s) = - \frac{J_{1} ^{2}}{4 \, J_{0} ^{2} \, c^{2}}
	\ee
	Therefore, the differential operator $\opO$ is given by
	\be
	\opO &\!\!=\!\!& \frac{\partial ^{2}}{\partial x ^{2}} - 2 \, \left( \frac{s}{c} + \frac{J_{1}}{2 \, J_{0} \, c} \right) \, \frac{\partial}{\partial x} + \frac{J_{1} ^{2}}{4 \, J_{0} ^{2} \, c^{2}} \, .
	\ee
	Then, the rescaled operator is easily computed, \ie
	\be
	\opL = \frac{\opO}{- 2 \, s / c} = \frac{\partial}{\partial x} + \frac{1}{s} \left[ - \frac{c}{2} \, \frac{\partial ^{2}}{\partial x ^{2}} +  \frac{J_{1}}{2 \, J_{0}} \, \frac{\partial}{\partial x} - \frac{J_{1} ^{2}}{8 \, c \, J_{0} ^{2}} \right] \, .
	\ee
	Hence, $\opL$ can be expressed in terms of decreasing powers of $s$ as follows,
	\be
	\opL \overset{s \to \infty}{\sim} \opL _{0} + \frac{1}{s} \, \opL _{1} \, ,
	\ee
	where
	\be
	\opL _{0} &\!\!=\!\!& \frac{\partial}{\partial x} \, , \\
	\opL _{1} &\!\!=\!\!& 
	- \frac{c}{2} \, \frac{\partial ^{2}}{\partial x ^{2}} +  
	\frac{J_{1}}{2 \, J_{0}} \, \frac{\partial}{\partial x} -
	 \frac{J_{1} ^{2}}{8 \, c \, J_{0} ^{2}} \, ,
	\ee
	from which we can also deduce that $\nu _{0} = 0$, $\nu _{1} = 1$ and $N = 1$.
	
	Now, the coefficients $\lambda _{k}$ are easily computed from the previous results, indeed 
	\be
	\lambda _{k} = m_{1} \, \nu _{1} = m_{1} \in \N \, 
	\ee
	thus, $\lambda _{k} = k$ for $k \in \N$. Moreover, from the condition
	$$ \nu _{i} + \lambda _{j} = \lambda _{k} \, ,  \qquad i = 1, 2 , \ldots \, ,$$
 we immediately infer that $j = k - \nu _{i}$.
 
 	Then, taking profit of the system in Eq.~(\ref{eq-system-A}), we can compute the values of the coefficients $A_{k, \ell}$ for the Maxwell model. In particular, given the previous discussion we have that the system in Eq.~(\ref{eq-system-A}) can be rewritten as
\begin{equation}
 \begin{cases}
  A_{k, 0} = \delta_{k 0} \qquad \ell=0 \\
  A_{k, \ell} = - \ds \sum _{i=1} \sum _{J} \left( p_i A_{J,\ell+1} + q_i A_{J, \ell} + r_i A_{J, \ell-1} \right) \delta_{J, \, k -  \nu _{i}} \qquad 1 \leq \ell \leq k \\
  A_{k, \ell} = 0 \qquad \ell>k
 \end{cases}
\end{equation}
In particular, the second line can be rewritten, dropping the sum and the delta, as
\begin{equation}
A_{k, \ell} = - \ds \Big( p_1 A_{k - \nu _{1}, \, \ell+1} + q_1 A_{k - \nu _{1}, \, \ell} + r_1 A_{k - \nu _{1}, \, \ell-1} \Big) \qquad 1 \leq \ell \leq k
\end{equation}
Thus, accordingly with the discussion in Section \ref{Sec-2}, we can immediately read off the values of all $p_{i}$'s, $q_{i}$'s and $r_{i}$'s from the operator $\mathbb{L} _{1}$, indeed
\begin{equation}
p_{1} = - \ds\frac{c}{2} \, , \,\,\, q_1= \ds\frac{J_{1}}{2\, J_{0}} \, , \,\,\, r_1= - \ds\frac{J_{1} ^{2}}{8 \, c \, J_{0} ^{2}} \, ,
\end{equation}
whereas $p_{i} = q_{i} = r_{i} = 0 \, , \,\, \forall i \geq 2$.	
	
	If we focus on the coefficients such that $1 \leq l \leq k$ we get
\be
A_{k, \ell} = - \ds \Big( - \ds\frac{c}{2} \, A_{k - 1, \, \ell+1} + \ds\frac{J_{1}}{2\, J_{0}} \, A_{k - 1, \, \ell} - \ds\frac{J_{1} ^{2}}{8 \, c \, J_{0} ^{2}} \, A_{k - 1, \, \ell-1} \Big)
\ee	
Thus,
\be \label{eq-Akl-max}
\boxed{A_{k, \ell} = - \frac{1}{2} \ds \Big( - c \, A_{k - 1, \, \ell+1} + \ds\frac{J_{1}}{J_{0}} \, A_{k - 1, \, \ell} - \ds\frac{J_{1} ^{2}}{4 \, c \, J_{0} ^{2}} \, A_{k - 1, \, \ell-1} \Big)} \, .
\ee	

	Considering an initial step input (\ie $r_{0} (s) = 1/s$), then the function $\wt{\Phi} _{k} (s, x)$ is given by
	\be
	\begin{split}
	\wt{\Phi} _{k} (s, x) &= s^{- (\lambda _{k} +1)} \, \exp \left[ - x \left( \mu _{+} (s) - \frac{s}{c} \right) \right] =\\
	&= \frac{1}{s ^{k + 1}} \, \exp \left( - \frac{x \, J_{1}}{2 \, J_{0} \, c} \right) \, ,
	\end{split}
	\ee 
	that, inverting it back to the time domain, gives
	\be
	\boxed{\Phi _{k} (t, x) = \exp \left( - \frac{x \, J_{1}}{2 \, J_{0} \, c} \right) \, \frac{t^{k}}{k!}} \, .
	\ee

	Given the previous results, we are now capable to write down the explicit wave-front expansion for the Maxwell model, \ie
	\begin{equation}
	r _{M} (t, x) \sim 
	\exp \left( - \frac{x \, J_{1}}{2 \, J_{0} \, c} \right) \,
	\sum _{k=0} ^\infty \,  \sum _{\ell = 0} ^k A_{k, \ell} \, \frac{x^\ell}{\ell !}  \, 
	\frac{(t - x)^{k}}{k!}  \, , \quad \mbox{as} \,\, t \to \left( \frac{x}{c} \right) ^{+} \, .
	\end{equation}
	where  the coefficients $A_{k, \ell}$ are determined as in Eq.~(\ref{eq-Akl-max}).

%%%%
%%%% Long Time Maxwell
\subsection{Long time asymptotic expansion}
	From Eq.~(\ref{eq-mu-maxwell}) one can easily deduce that
	\be
	\mu (s) \overset{s \to 0}{\sim} \sqrt{\rho \, J_{1}} \, s^{1/2} \, ,
	\ee
	thus,
	\be
	\wt{r} _{M} (s, x) = \frac{1}{s} \, \exp \left[ - x \, \mu (s) \right] \overset{s \to 0}{\sim} 
	\frac{1}{s} \, \exp \left( - x \, \sqrt{\rho \, J_{1}} \, s^{1/2} \right) \, .
	\ee
Now, inverting back to the time domain we get
\be
\boxed{r_{M} (t, x) \overset{t \to \infty}{\sim} \texttt{erfc} \left( \frac{1}{2} \, \sqrt{\frac{x^{2} \, J_{1} \, \rho}{t}} \right)} \, .
\ee
where \texttt{erfc} stands for the complementary error function.

\subsection{Numerical Results} 
	In the following we show separately the plots of both the asymptotic expansions for the ordinary Maxwell model, at fixed values of $x$. Then, we highlight an explicit matching between the wave-front expansion and the long time expansion, setting $\rho=J_0=J_1=1$.

\begin{figure}[h!]
\centering
\includegraphics[width=13cm]{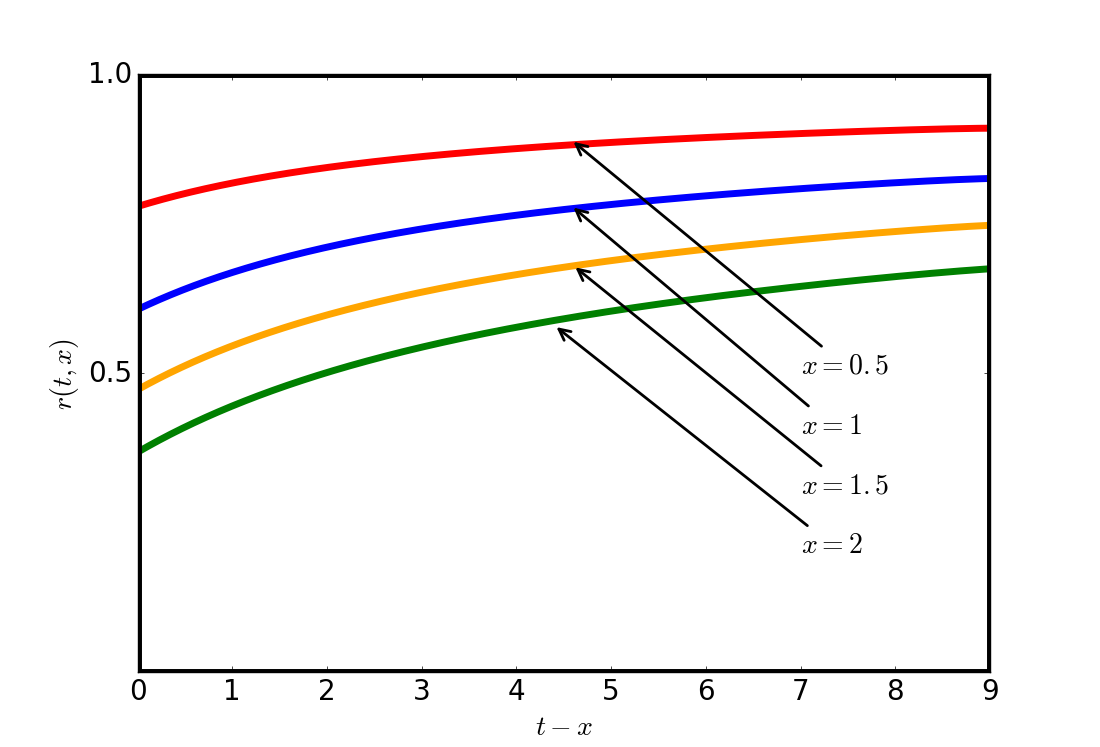}
\caption{The wave-front expansion for the ordinary Maxwell model. Clearly, this approximation cannot be trusted for values of $t - x$ for which the expansion loses its monotonic behavior.}
\end{figure}	

\begin{figure}[h!]
\centering
\includegraphics[width=13cm]{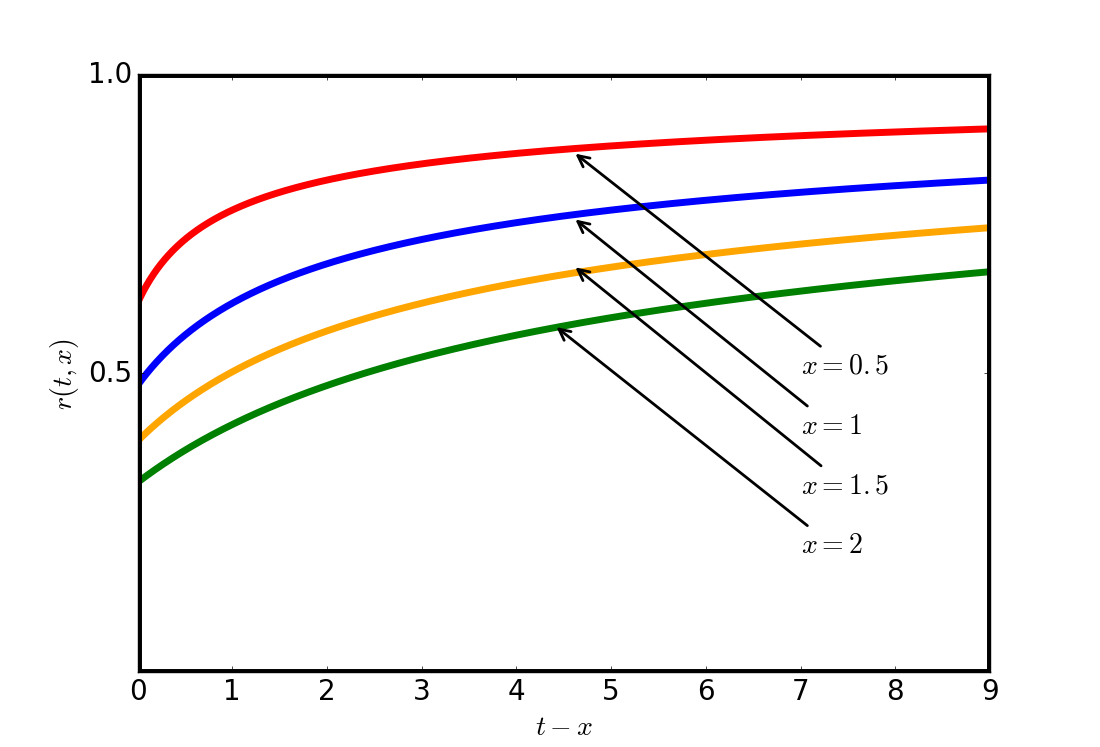}
\caption{The long time asymptotic expansion for the ordinary Maxwell model.}
\end{figure}		

\begin{figure}[h!]
\centering
\includegraphics[width=13cm]{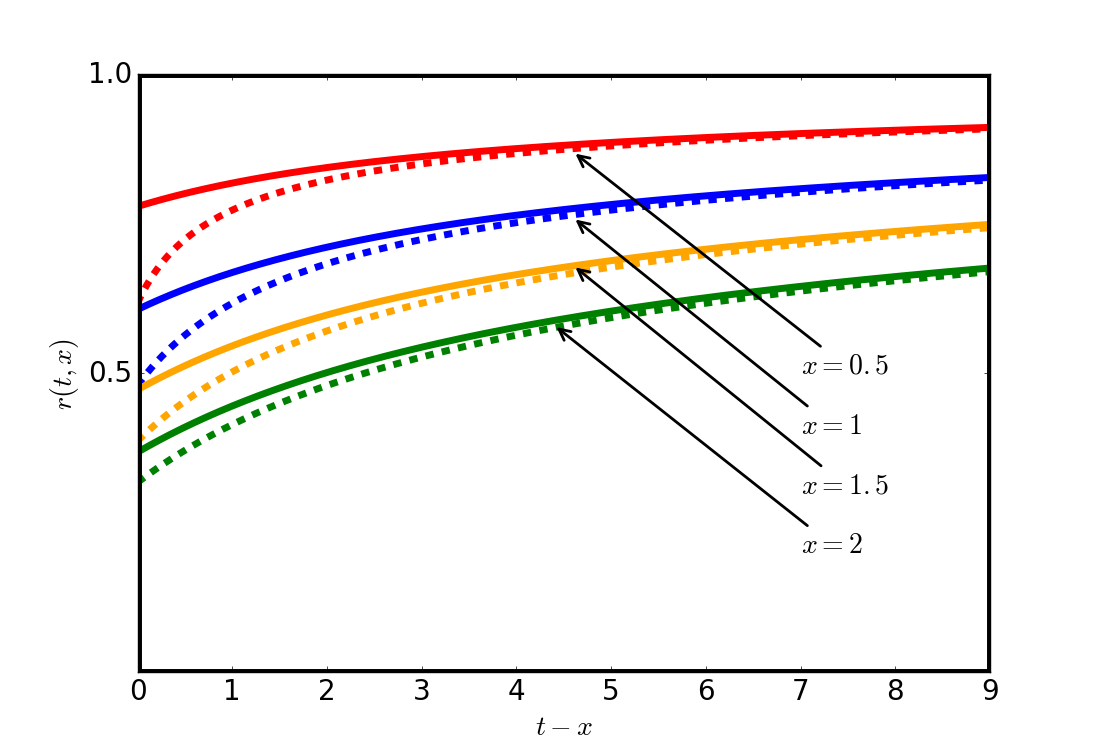}
\caption{The matching between the two asymptotic expansions for the ordinary Maxwell model. The solid lines represent the wave-front expansion, and the dashed lines represent the long time expansion.}
\end{figure}

%%%%%%%%%% GENERAL FRACTIONAL MAXWELL
\newpage
\section{The Fractional Maxwell Model of order $\alpha$} \label{Sec-4}

Similarly to the ordinary case, the fractional Maxwell model of linear viscoelasticity is defined by means of its constitutive equation
\be
\sigma (t) + a_1 \, D^\alpha _t \sigma (t) = b_1 \, D^\alpha _t \varepsilon (t)
\ee
where the constants $a_{1}$ and $b_{1}$ are required to be strictly positive. 

	Here, the time derivatives are replaced by the well known fractional Caputo derivative of order $\alpha \in \R^+$ (see \cite{Mainardi_BOOK10}), defined, for a certain function of time $f(t)$, as
\be \label{eq-frac-der}
D^\alpha _t f(t) = 
\left\{
\begin{aligned}
 \frac{1}{\Gamma (m-\alpha)} \int _0 ^t d\tau \frac{f^{(m)} (\tau)}{(t-\tau)^{\alpha +1-m}} \,,  \qquad & m-1 < \alpha <m \\
 \frac{d^m}{dt ^m} f(t) \,, \qquad \qquad \qquad \qquad \qquad & \alpha = m
\end{aligned}
\right.
\,,
\ee
that, assuming appropriate initial condition of physical interest for viscoelastic materials, can be replaced as follows when the Laplace transform is applied to both sides of a constitutive equation
\be
D^\alpha _t f(t) \div s ^\alpha \, \wt f (s) \,,
\ee
for further details on the initial conditions we invite the reader to refer to \cite{AG-FCAA-2017, Gurtin-Sternberg, Mainardi_BOOK10}.

	Then we can finally infer that
\be \label{eq:sJ-fracmax}
s \, \wt{J} _{M, \alpha} (s) = \frac{ a_{1}}{b_{1} } \left[ 1+	\frac{1}{(s\, \tau)^\alpha} \right] \, ,
\ee	
where $\tau ^\alpha = a_1$.

\subsection{Wave-front expansion for the general case}

For the fractional Maxwell case we find
\be
\mu (s) = \sqrt{\rho} \, s	\, \left[s \, \wt{J} _{M, \alpha} (s) \right] ^{1/2} =
\sqrt{\frac{\rho\, a_{1}}{b_{1}}} \,s\, \left[
1+	\frac{1}{(s\, \tau)^\alpha}	\right]^{1/2} \,.
\ee	

	Again, as $s \to \infty$ one can easily find that
	\be
	\mu (s) &\!\!=\!\!& \sqrt{\frac{\rho\, a_{1}}{b_{1}}} \,s\, \left[
1+	\frac{1}{(s\, \tau)^\alpha}	\right]^{1/2} = \frac{1}{c} \,s\, \left[ 1+\sum_{n=1}^{\infty} \binom {1/2} {n} \frac{1}{(s\, \tau)^{\alpha\, n}} \right]=\\
	&\!\!=\!\!&
	\frac{1}{c} \left[
	s + \sum_{n=1}^{\infty}  \binom {1/2} {n}\,
	\frac{s^{1-\alpha\, n}}{a_1^n}
	  \right] \notag
	\ee	
where we have taken profit of the relations $c = \sqrt{b_1/(\rho \, a_1)}$ and $\tau^\alpha = a_1$.

At this point, since $\mu_+ (s)$ has to contain only positive powers of $s$, it is recommended to distinguish between different cases. As $1-n\alpha \geq 0$, then we have to satisfy the condition $n\leq [1/\alpha]$. So, if $\alpha=1/2 $, we find that $n=1,2$, if $\alpha>1/2 $, $n=1$ and if $\alpha<1/2 $ there are several values of $n$ (\eg $\alpha=1/4$ implies $n=1,2,3,4$).

\subsection{The Fractional Maxwell Model of order 1/2}
If we fix $\alpha = 1/2$, we have that the \eqref{eq:sJ-fracmax} becomes
\be
s \, \wt{J} _{M, 1/2} (s) = \frac{ a_{1}}{b_{1} } \left[ 1+	\frac{1}{(s\, \tau)^{1/2}} \right] \, .
\ee	

\subsubsection{Wave-front Expansion}

As a consequence, we can perform the wave-front expansion, and we have
\be \label{eq-mu-fracmax1/2}
\mu (s) = 
\sqrt{\frac{\rho\, a_{1}}{b_{1}}} \,s\, \left[
1+	\frac{1}{(s\, \tau)^{1/2}}	\right]^{1/2} \,.
\ee	
From the previous general case, as $s \to \infty$ we find that
	\be
	\mu (s) &\!\!=\!\!& 
	\frac{1}{c}\left[
	s + \frac{s^{1/2}}{2a_1} -\frac{1}{8a_1^2}+
	\sum_{n=3}^{\infty}  \binom {1/2} {n}\,
	\frac{s^{1- n/2}}{a_1^n}
	  \right] \,,
	\ee	
and, from the latter, we can easily read
	\be
	\mu _{+} (s) &\!\!=\!\!& 
	\frac{1}{c}\left[
	s + \frac{s^{1/2}}{2a_1} -\frac{1}{8a_1^2} \right]
	 \, , \\
	\mu _{-} (s) &\!\!=\!\!& \frac{1}{c} 
	\sum_{n=3}^{\infty} \binom {1/2} {n}\,
	\frac{s^{1- n/2}}{a_1^n} \, .
	\ee
	Then,
	\be
	\mu ^{2} (s) &\!\!=\!\!& \frac{s^{2}}{c^{2}} \left[ 1+\frac{1}{a_1\,s^{1/2}} \right]=\frac{1}{c^2}\left[ s^2 + \frac{s^{3/2}}{a_1} \right] \, , \\
	\mu _{+} ^{2} (s) &\!\!=\!\!& \frac{1}{c^{2}}\left[ s^2 + \frac{s^{3/2}}{a_1} -\frac{s^{1/2}}{8\,a_1^3} + \frac{1}{64\,a_1^4} \right] \, ,
	\ee
	from which we get
	\be
	\mu ^{2} (s) - \mu _{+} ^{2} (s) = \frac{1}{c^2} \left[ \frac{s^{1/2}}{8\,a_1^3}-\frac{1}{64\,a_1^4} \right] \, .
	\ee

	Therefore, the differential operator $\opO$, defined in \eqref{eq-L}, is given by
	\be
	\opO &\!\!=\!\!& \frac{\partial ^{2}}{\partial x ^{2}} - \frac{2}{c} \, \left( s + \frac{s^{1/2}}{2a_1} -\frac{1}{8a_1^2} \right) \, \frac{\partial}{\partial x} - \frac{1}{c^2} \left[ \frac{s^{1/2}}{8\,a_1^3}-\frac{1}{64\,a_1^4} \right] \, .
	\ee
	At this point, the rescaled operator is easily computed, \ie
	\be
	\opL &\!\!=\!\!& \frac{\opO}{- 2 \, s / c} =\\
	&\!\!=\!\!&
	 \frac{\partial}{\partial x} +
\frac{1}{s^{1/2}}\left[ \frac{1}{2\,a_1} \frac{\partial}{\partial x} + \frac{1}{16\, c\, a_1^3} \right] +
	 \frac{1}{s} \left[ - \frac{c}{2} \, \frac{\partial ^{2}}{\partial x ^{2}} -  \frac{1}{8 \, a_{1}^2} \, \frac{\partial}{\partial x} - \frac{1}{128 \, c \, a_{1} ^{4}} \right] \notag \, .
	\ee
	Hence, $\opL$ can be expressed in terms of decreasing powers of $s$ as follows,
	\be
	\opL \overset{s \to \infty}{\sim} \opL _{0} + \frac{1}{s^{1/2}} \, \opL _{1}  +\frac{1}{s} \, \opL _{2} \, ,
	\ee
	where
	\be
	\opL _{0} &\!\!=\!\!& \frac{\partial}{\partial x} \, , \\
	\opL _{1} &\!\!=\!\!& 
	\frac{1}{2\,a_1} \frac{\partial}{\partial x} + \frac{1}{16\, c\, a_1^3}\\
	\opL _{2} &\!\!=\!\!&
	- \frac{c}{2} \, \frac{\partial ^{2}}{\partial x ^{2}} -  \frac{1}{8 \, a_{1}^2} \, \frac{\partial}{\partial x} - \frac{1}{128 \, c \, a_{1} ^{4}}	
	 \, ,
	\ee
	and, from which we can also deduce that $\nu _{0} = 0$, $\nu _{1} = 1/2$, $\nu _{2} = 1$ and $N = 2$.

	Now, the coefficients $\lambda _{k}$ are easily computed taking into account the previous results, indeed 
	\be
	\lambda _{k} = m_{1} \, \nu _{1} + m_{2} \, \nu _{2} = \frac{m_1}{2} + m_2 \,, \qquad m_{1}, m_{2} \in \N \, 
	\ee
	thus, $\lambda _{k} = k/2$ for $k \in \N$. Moreover, from the condition
	$$ \nu _{i} + \lambda _{j} = \lambda _{k} \, ,  \qquad i = 1, 2 , \ldots \, ,$$
 we immediately infer that $j = k - 2 \, \nu _{i}$.

Thus, following the algorithm in Section \ref{Sec-2}, we can immediately read off the values of all $p_{i}$'s, $q_{i}$'s and $r_{i}$'s from the expressions for $\mathbb{L} _{1}$ and $\opL _{2}$, indeed
\be
p_{1} = 0 \, , \,\,\, q_1= \ds\frac{1}{2\,a_1} \, , \,\,\, r_1= \ds\frac{1}{16 \, c \, a_1 ^{3}} \, ,\\
p_{2} = - \ds\frac{c}{2} \, , \,\,\, q_2= -\ds\frac{1}{8\, a_1^2} \, , \,\,\, r_2= - \ds\frac{1}{128 \, c \, a_{1} ^{4}}
\ee
while $p_{i} = q_{i} = r_{i} = 0 \, , \,\, \forall i \geq 3$.	
	
	Finally, if we focus on the coefficients such that $1 \leq l \leq k$ we obtain the relation
\begin{align} \label{eq-Akl-frac_max}
A_{k, \ell} &= - \frac{1}{2} \ds \Big( \ds\frac{1}{a_{1}} \, A_{k - 1, \, \ell} + \ds\frac{1}{8 \, c \, a_{1} ^{3}} \, A_{k - 1, \, \ell-1} + \notag\\
&- c\, A_{k - 2, \, \ell+1} -\frac{1}{4\,a_1^2}\, A_{k - 2, \, \ell} -\frac{1}{64\, c\, a_1^4} \, A_{k - 2, \, \ell-1} \Big) \, .
\end{align}	

Considering an initial step input (\ie $r_{0} (s) = 1/s$), then the function $\wt{\Phi} _{k} (s, x)$ is given by
	\be
	\begin{split}
	\wt{\Phi} _{k} (s, x) &= s^{- (\lambda _{k} +1)} \, \exp \left[ - x \left(\mu _{+} (s) -\frac{s}{c}\right) \right] =\\
	&= \frac{1}{s ^{\frac{k}{2} + 1}} \, \exp \left[ - x\, \left( \frac{1}{2ca_1} \, s^{1/2} - \frac{1}{8\,c\,a_1^2} \right) \right] \, ,
	\end{split}
	\ee 
	that, inverting it back to the time domain, gives
	\be
	\Phi _{k} (t, x) = \exp\left( \frac{x}{8\,c\,a_1^2} \right) t^{k/2} \, F_{1/2} \left( \frac{x}{2\,c\,a_1\sqrt{t}}, \frac{k}{2} \right) \, .
	\ee
	Hence, given the previous results, the wave-front expansion for the fractional Maxwell model of order $1/2$ is given by
	\begin{equation}
	r _{M,1/2} (t, x) \sim \exp\left( \frac{x}{8\,c\,a_1^2} \right)
	\sum _{k=0} ^\infty \,  \sum _{\ell = 0} ^k A_{k, \ell} \, \frac{x^\ell}{\ell !}  \, 
	(t-x)^{k/2} \, F_{1/2} \left(\frac{x}{2\,c\,a_1\sqrt{t-x}}, \frac{k}{2} \right)
	 \, , 
	\end{equation}
as $t \to \left( x / c \right) ^{+}$.

\subsubsection{Long time asymptotic expansion}
	From Eq.~(\ref{eq-mu-fracmax1/2}) one can easily deduce that
	\be
	\mu (s) \overset{s \to 0}{\sim} \sqrt{\frac{\rho}{b_1}} s^{3/4} \, ,
	\ee
	thus,
	\be
	\wt{r} _{M, 1/2} (s, x) = \frac{1}{s} \, \exp \left[ - x \, \mu (s) \right] \overset{s \to 0}{\sim} 
	\frac{1}{s} \, \exp \left( - x \, \sqrt{\rho / b_{1}} \, s^{3/4} \right) \, .
	\ee
Thereafter, inverting the latter expression back to the time domain we get \cite{extra2}
\be
\boxed{r_{M, 1/2} (t, x) \overset{t \to \infty}{\sim} W_{-3/4, 1} \left( - \sqrt{\frac{\rho}{b_1}} \frac{x}{t^{3/4}} \right)} \, ,
\ee
where $W_{ \gamma,\delta }$ represents the Wright function
\be
	W_{\gamma, \delta} (z) = \sum_{k=0}^\infty
	\frac{z^k}{\ds k! \, \Gamma\left(\gamma k+\delta  \right)}\,, \qquad \gamma > -1 \,, \qquad \delta \in  \mathbb{C}\, ,
\ee 
which satisfies
\be
	s^{-\delta} \, \exp{\left[-\lambda\, s^{-\gamma}\right]} \div t^{\delta -1} \, W_{\gamma, \delta} \left(-\lambda\, t^\gamma \right) \,,
	\qquad -1<\gamma <0 \,,\qquad \lambda>0 \,.
\ee
For further information about Wright functions see \cite{Gorenflo-Luchko-Mainardi}.

\subsubsection{Numerical Results} 
	In the following we show separately the plots of both the asymptotic expansions for the fractional Maxwell model of order $1/2$. Similarly to the ordinary Maxwell model, we consider plots at fixed position $x$. Then, we show an explicit matching between the wave-front expansion and the long time expansion, setting $a_1=b_1=\rho=1$ so that $c=1$.

\begin{figure}[h!]
\centering
\includegraphics[width=13cm]{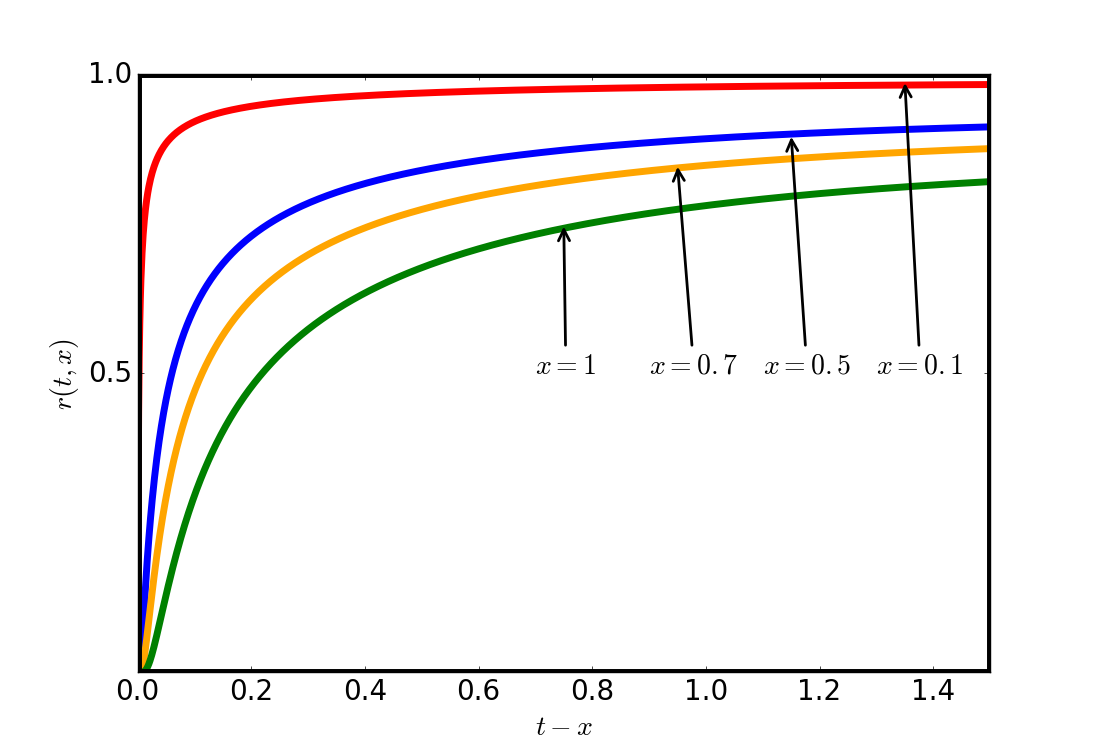}
\caption{The wave-front expansion for the fractional Maxwell model of order 1/2. Clearly, this approximation cannot be trusted for values of $t - x$ for which the expansion loses its monotonic behaviour.}
\end{figure}	

\begin{figure}[h!]
\centering
\includegraphics[width=13cm]{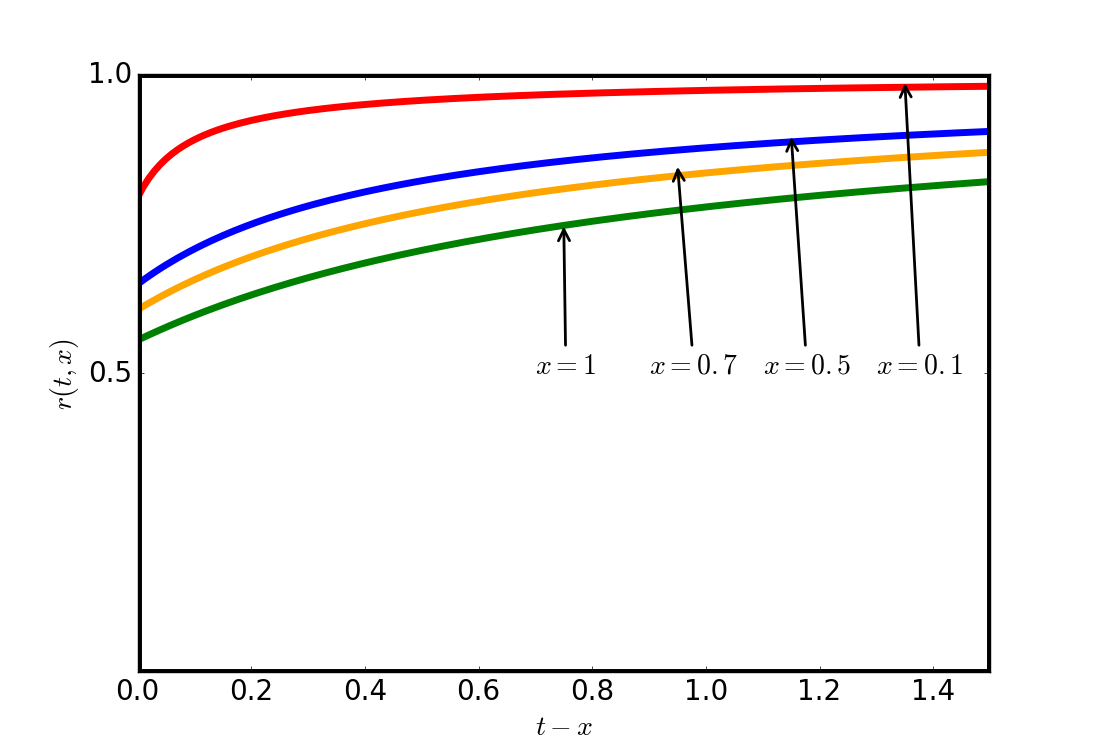}
\caption{The long time asymptotic expansion for the fractional Maxwell model of order $1/2$.}
\end{figure}		

\begin{figure}[h!]
\centering
\includegraphics[width=13cm]{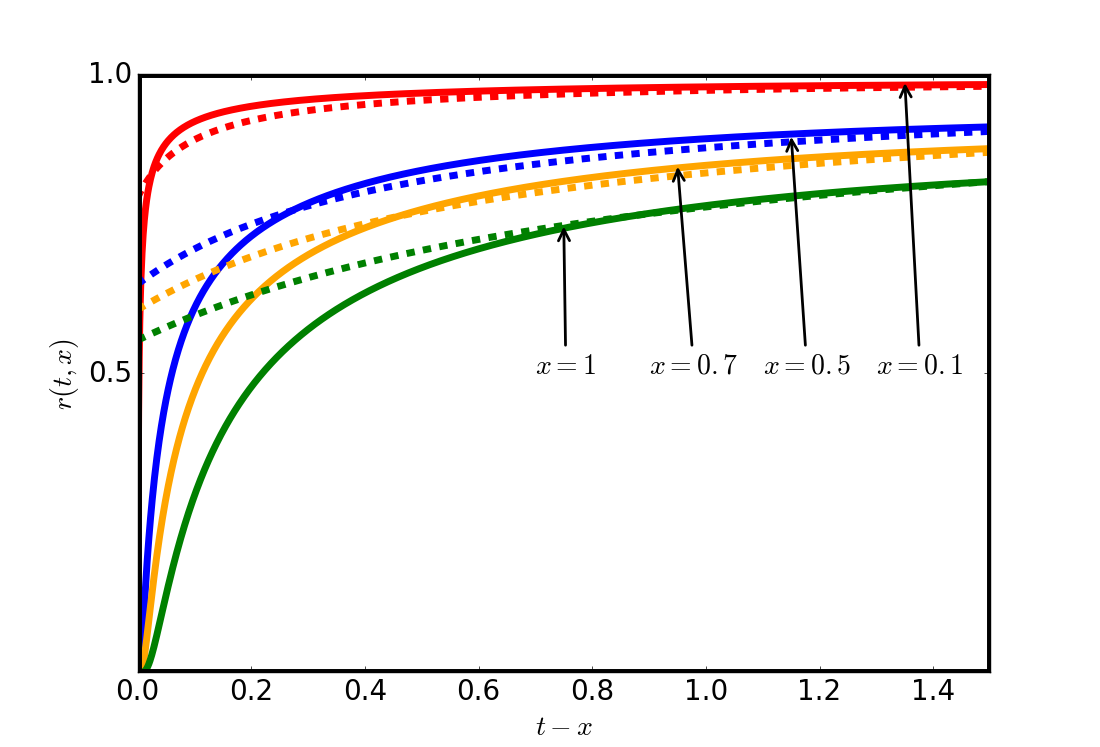}
\caption{The matching between the two asymptotic expansions for the fractional Maxwell model of order $1/2$. The solid lines represent the wave-front expansion, and the dashed lines represent the long time expansion.}
\end{figure}

%%%%%%%%%%%%%%%%%%%%%%%%%%%%%%%
%%%%%%% FRACTIONAL MAXWELL 3/4
%%%%%%%%%%%%%%%%%%%%%%%%%%%%%%%
\newpage
\subsection{The Fractional Maxwell Model of order 3/4}
Now we set $\alpha = 3/4$, so that the creep compliance in the Laplace domain reads
\be
s \, \wt{J} _{M, 3/4} (s) = \frac{ a_{1}}{b_{1} } \left[ 1+	\frac{1}{(s\, \tau)^{3/4}} \right] \, ,
\ee	
and than we repeat the procedure discussed above.

\subsubsection{Wave-front Expansion}
First write
\be \label{eq-mu-fracmax3/4}
\mu (s) = 
\sqrt{\frac{\rho\, a_{1}}{b_{1}}} \,s\, \left[
1+	\frac{1}{(s\, \tau)^{3/4}}	\right]^{1/2} \,.
\ee	
Secondly, we compute the limit for $s \to \infty$ of \eqref{eq-mu-fracmax3/4}, namely
	\be
	\mu (s) &\!\!=\!\!& 
	\frac{1}{c}\left[
	s + \frac{s^{1/4}}{2a_1} +
	\sum_{n=2}^{\infty}  \binom {1/2} {n}\,
	\frac{s^{1- 3n/4}}{a_1^n}
	  \right] \,.
	\ee	
	Then, from the previous expression, we immediately read off
	\be
	\mu _{+} (s) &\!\!=\!\!& 
	\frac{1}{c}\left[
	s + \frac{s^{1/4}}{2a_1} \right]
	 \, , \\
	\mu _{-} (s) &\!\!=\!\!& \frac{1}{c} 
	\sum_{n=2}^{\infty} \binom {1/2} {n}\,
	\frac{s^{1- 3n/4}}{a_1^n} \, ,
	\ee
	also inferring that
	\be
	\mu ^{2} (s) &\!\!=\!\!& \frac{s^{2}}{c^{2}} \left[ 1+\frac{1}{a_1\,s^{3/4}} \right]=\frac{1}{c^2}\left[ s^2 + \frac{s^{5/4}}{a_1} \right] \, , \\
	\mu _{+} ^{2} (s) &\!\!=\!\!& \frac{1}{c^{2}}\left[ s^2 + \frac{s^{5/4}}{a_1} + \frac{s^{1/2}}{4\,a_1^2}\right] \, ,
	\ee
	from which we get
	\be
	\mu ^{2} (s) - \mu _{+} ^{2} (s) = -\frac{s^{1/2}}{4\,c^2\,a_1^2} \, .
	\ee

	Hence, the differential operator $\opO$ is given by
	\be
	\opO &\!\!=\!\!& \frac{\partial ^{2}}{\partial x ^{2}} - \frac{2}{c} \, \left( s + \frac{s^{1/4}}{2a_1}\right) \, \frac{\partial}{\partial x} + \frac{s^{1/2}}{4\,c^2\,a_1^2} \, ,
	\ee
	that can be rescaled as
	\be
	\opL &\!\!=\!\!& \frac{\opO}{- 2 \, s / c} =
	\frac{\partial}{\partial x} -
\frac{1}{8\,c\,a1^2\,s^{1/2}} + \frac{1}{2a_1\,s^{3/4}}\,\frac{\partial}{\partial x} - \frac{c}{2s} \, \frac{\partial ^{2}}{\partial x ^{2}} \, .
	\ee
	At this point, $\opL$ can be expressed in terms of decreasing powers of $s$ as follows,
	\be
	\opL \overset{s \to \infty}{\sim} \opL _{0} + \frac{1}{s^{1/2}} \, \opL _{1}  +\frac{1}{s} \, \opL _{2} \, ,
	\ee
	where
	\be
	\opL _{0} &\!\!=\!\!& \frac{\partial}{\partial x} \, , \\
	\opL _{1} &\!\!=\!\!& 
	-\frac{1}{8\,c\,a_1^2}\\
	\opL _{2} &\!\!=\!\!& 
	-\frac{1}{2\,a_1}\, \frac{\partial}{\partial x}\\
	\opL _{3} &\!\!=\!\!&
	- \frac{c}{2} \, \frac{\partial ^{2}}{\partial x ^{2}} 
	 \, ,
	\ee
	and from these expansions we immediately deduce that $\nu _{0} = 0$, $\nu _{1} = 1/2$, $\nu _{2} = 3/4$, $\nu _{3} = 1$ and $N = 3$.

	Given that, the coefficients $\lambda _{k}$ are easily computed from the previous results, so that 
	\be
	\lambda _{k} = m_{1} \, \nu _{1} + m_{2} \, \nu _{2} = \frac{m_1}{2} + \frac{3m_2}{4}+ m_3 \,, \qquad m_{1}, m_{2}, m_{3} \in \N \, 
	\ee
	accordingly, $\lambda _{k} = k/4$ for $k \in \N$. Moreover, considering the condition
	$$ \nu _{i} + \lambda _{j} = \lambda _{k} \, ,  \qquad i = 1, 2 , \ldots \, ,$$
 we conclude that $j = k - 4 \, \nu _{i}$.

Thus, following the procedure described by the algorithm, we can directly realise the values of all the coefficients $p_{i}$'s, $q_{i}$'s and $r_{i}$'s from the expressions for every $\opL _{i}$, indeed
\be
p_{1} = 0 \, , \,\,\, q_1= 0 \, , \,\,\, r_1= \ds-\frac{1}{8 \, c \, a_1 ^{2}} \, ,\\
p_{2} = 0 \, , \,\,\, q_2= \ds\frac{1}{2\, a_1} \, , \,\,\, r_2= 0 \, , \\
p_{3} = \ds -\frac{c}{2} \, , \,\,\, q_3= 0 \, , \,\,\, r_3= 0 \, ,
\ee
while $p_{i} = q_{i} = r_{i} = 0 \, , \,\, \forall i \geq 4$.	
	
	Focusing our attention on the coefficients such that $1 \leq l \leq k$, we get
\begin{align} \label{eq-Akl-frac_max3/4}
A_{k, \ell} &= - \frac{1}{2} \ds \Big( \ds\frac{1}{4\,c\,a_{1}^2} \, A_{k - 2, \, \ell -1} + \ds\frac{1}{a_{1}} \, A_{k - 3, \, \ell} - c\, A_{k - 4, \, \ell+1} \Big) \, .
\end{align}	

	Now, we consider an initial step input (\ie $r_{0} (s) = 1/s$), so that the function $\wt{\Phi} _{k} (s, x)$ is given by
	\be
	\begin{split}
	\wt{\Phi} _{k} (s, x) &= s^{- (\lambda _{k} +1)} \, \exp \left[ - x \left(\mu _{+} (s) -\frac{s}{c}\right) \right] =\\
	&= \frac{1}{s ^{\frac{k}{4} + 1}} \, \exp \left[ -\frac{x}{2\,c\,a_1} s^{1/4} \right] \, ,
	\end{split}
	\ee 
	that, inverting back to the time domain, gives
	\be
	\Phi _{k} (t, x) =  t^{k/4} \, W_{-1/4, k/4 -1} \left( -\frac{x}{2\,c\,a_1\, t^{1/4}} \right) \, .
	\ee
	
	Then, from Eq. \eqref{eq-fin} we have that the wave-front expansion for the fractional Maxwell model of order $3/4$ is given by
	\begin{equation}
	r _{M,3/4} (t, x) \sim 
	\sum _{k=0} ^\infty \,  \sum _{\ell = 0} ^k A_{k, \ell} \, \frac{x^\ell}{\ell !}  \, 
	(t-x)^{k/4} W_{-1/4, k/4 -1} \left( -\frac{x}{2\,c\,a_1\, (t-x)^{1/4}} \right)
	 \, , 
	\end{equation}
as one approaches the wave-front. It is important to recall that $W_{ \gamma,\delta }$ represents the Wright function introduced above for the fractional Maxwell of order $1/2$.

\subsubsection{Long time asymptotic expansion}
	From Eq.~(\ref{eq-mu-fracmax3/4}) one can easily deduce that
	\be
	\mu (s) \overset{s \to 0}{\sim} \sqrt{\frac{\rho}{b_1}} s^{5/8} \, ,
	\ee
	so that,
	\be
	\wt{r} _{M, 3/4} (s, x) = \frac{1}{s} \, \exp \left[ - x \, \mu (s) \right] \overset{s \to 0}{\sim} 
	\frac{1}{s} \, \exp \left( -\frac{x}{c\,a_1^{1/2}} \, s^{5/8} \right) \, .
	\ee
	Now, inverting the latter back to the time domain, we finally get
\be
\boxed{r_{M, 3/4} (t, x) \overset{t \to \infty}{\sim} W_{-5/8, 1} \left( -  \frac{x}{c\,a_1^{1/2}\,t^{5/8}} \right)} \, .
\ee

\subsubsection{Numerical Results} 
	In the following we show separately the plots of both the asymptotic expansions for the fractional Maxwell model of order $3/4$. Then, we show an explicit matching between the wave-front expansion and the long time expansion, setting $a_1=b_1=\rho=1$ so that $c=1$.

\begin{figure}[h!]
\centering
\includegraphics[width=13cm]{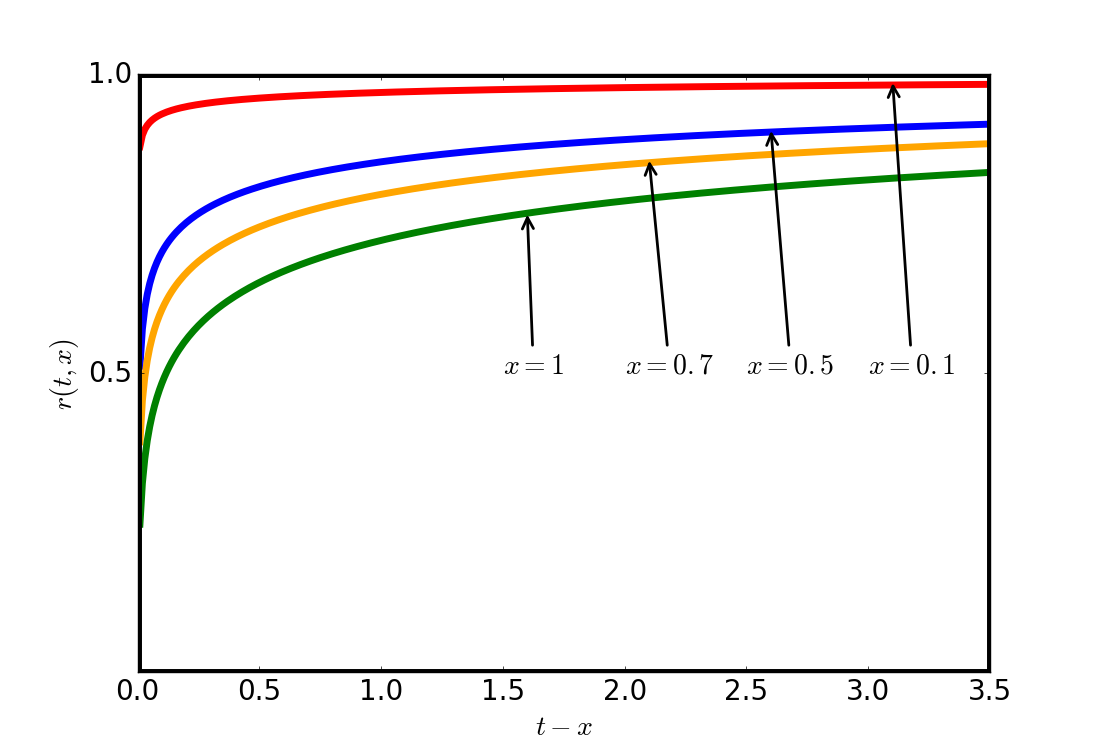}
\caption{The wave-front expansion for the fractional Maxwell model of order 3/4. Clearly, this approximation cannot be trusted for values of $t - x$ for which the expansion loses its monotonic behaviour.}
\end{figure}

\begin{figure}[h!]
\centering
\includegraphics[width=13cm]{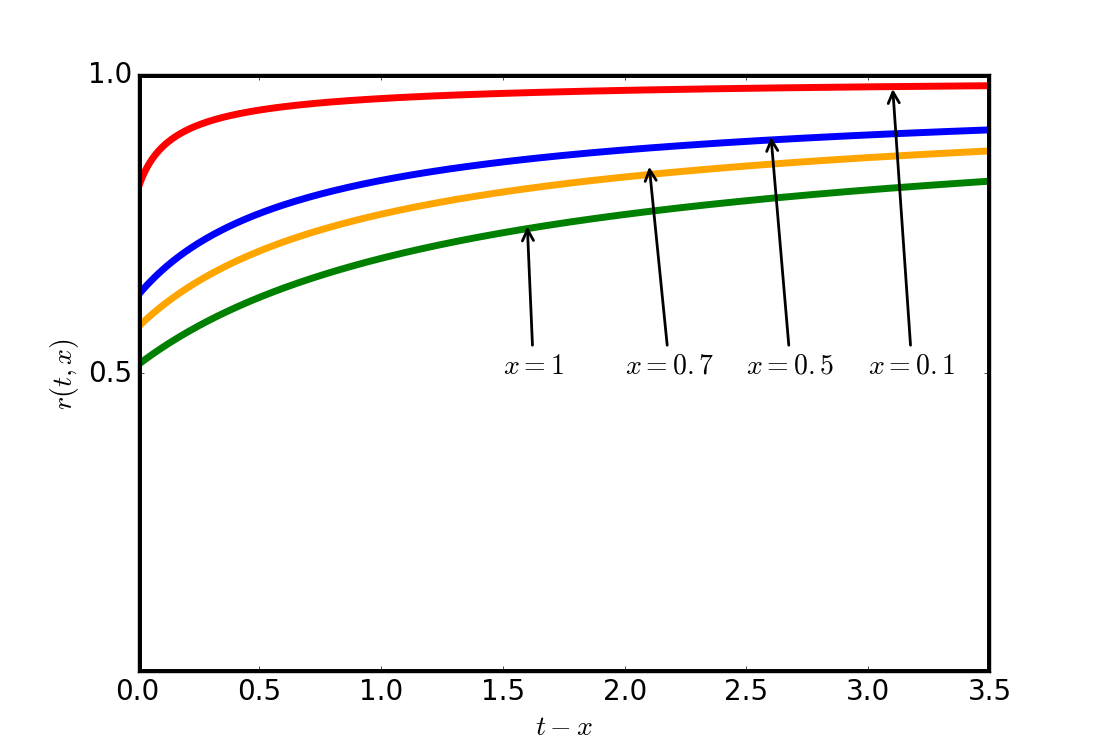}
\caption{The long time asymptotic expansion for the fractional Maxwell model of order 3/4.}
\end{figure}		

\begin{figure}[h!]
\centering
\includegraphics[width=13cm]{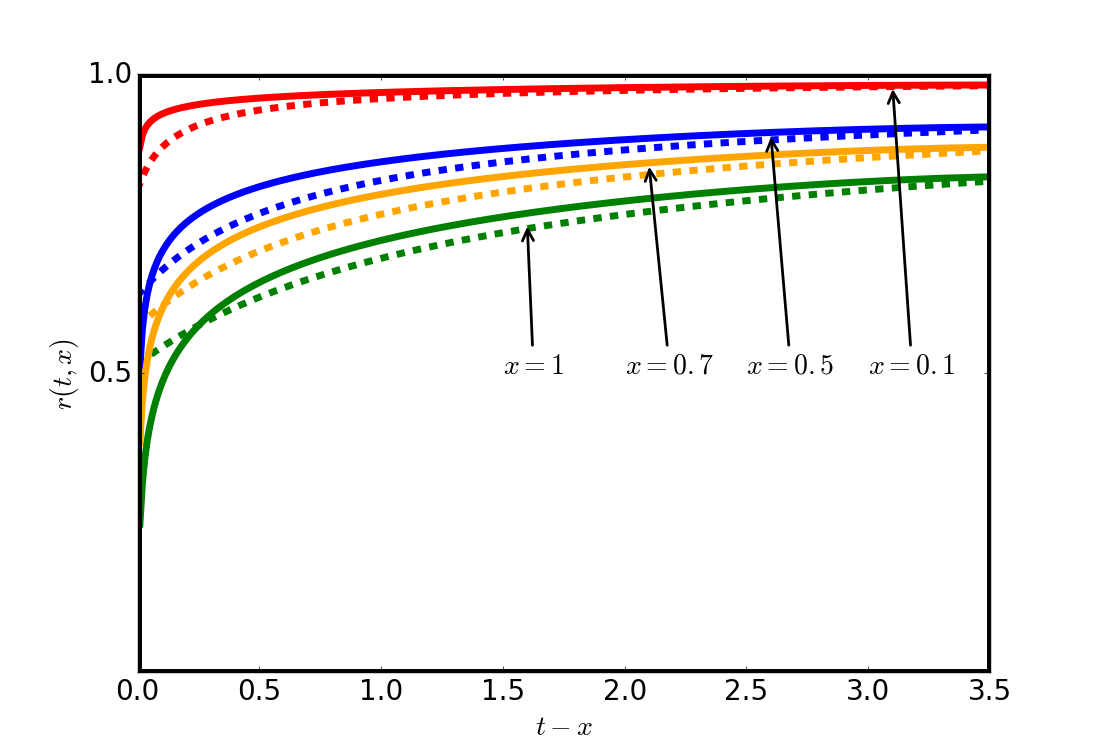}
\caption{The matching between the two asymptotic expansions for the fractional Maxwell model for $\alpha= 3/4$. The solid lines represent the wave-front expansion, and the dashed lines represent the long time expansion.}
\end{figure}

%%%%%%% ORDINARY VOIGT %%%%%%%
%%%%%%%%%%%%%%%%%%%%%%%%

\section{The (Ordinary) Voigt Model} \label{Sec-5}
	The ordinary Voigt model is defined in terms of its constitutive equation that reads
	\be
	\sigma (t) = m \, \varepsilon (t) + b_{1} \, \frac{d \varepsilon (t)}{dt} \, ,
	\ee
where $m \geq 0$ and $b_{1}$ is a strictly positive constant.
	
	In the Laplace domain the latter is rewritten as
	\be
	\wt{\sigma} (s) = ( m + b_{1} \, s) \, \wt{\varepsilon} (s) \, ,
	\ee
from which we can immediately read off the Laplace transform of the creep compliance,
	\be \label{eq-J-voigt}
	\wt{J} _{V} (s) = \frac{J_{1}}{s \, (1 + \tau _{\varepsilon} \, s)} \, ,
 	\ee
where $J_{1} = 1/m$ and $\tau _{\varepsilon} = b_{1} / m$.
	
	For sake of completeness, the creep compliance defined in Eq.~(\ref{eq-J-voigt}) can be inverted back to the time domain and it gives
	\be
	J_{V} (t) = J_{1} \, \left[ 1 - \exp \left(- \frac{t}{\tau _{\varepsilon}} \right) \right] \, ,
	\ee
	from which one can infer that the wave-front velocity is infinite (\ie $J_{0} = 0$).
%%%
	\subsection{Wave-front Expansion}
	Here we have that
	\be \label{eq-mu-voigt}
	\mu (s) = \sqrt{\rho} \, s \, \left[s \, \wt{J} _{V} (s) \right] ^{1/2} 		= \sqrt{\rho} \, s \, \left[\frac{J_{1}}{1 + \tau _{\varepsilon} \, s} \right] ^{1/2} \,,
	\ee
and from asymptotic behaviour for $s \to \infty$ we have
	\be	
	\mu (s) \overset{s \to \infty}{\sim}
	\sqrt{\frac{\rho \, J_{1}}{\tau _{\varepsilon}}} \, s^{1/2} \, \left(1 + o[(s \, \tau _{\varepsilon}) ^{-1}] \right) \,.
	\ee
Hence, it is straightforward to see that
	\be
	\mu _{+} (s) = \sqrt{\frac{\rho \, J_{1}}{\tau _{\varepsilon}}} \, s^{1/2} \, .
	\ee
Moreover, we have that
	\be
	\mu ^{2} (s) &\!\!=\!\!& \rho \, J_{1} \, \frac{s^{2}}{1 + \tau _{\varepsilon} \, s} \, , \\
	\mu ^{2} _{+} (s) &\!\!=\!\!& \frac{\rho \, J_{1}}{\tau _{\varepsilon}} \, s \, .
	\ee

	Consequently, the last expressions tell us that
	\be
	\mu ^{2} (s) - \mu ^{2} _{+} (s) = 
	- \frac{\rho \, J_{1}}{\tau _{\varepsilon}} \, \frac{s}{1 + \tau _{\varepsilon} \, s} \, .
	\ee
	
Now, the differential operator $\opO$ for the ordinary Voigt model is given by
\be
\opO &\!\!=\!\!& \frac{\partial ^{2}}{\partial x ^{2}} - 2 \, \left( \sqrt{\frac{\rho \, J_{1}}{\tau _{\varepsilon}}} \, s^{1/2} \right) \, \frac{\partial}{\partial x} + \frac{\rho \, J_{1}}{\tau _{\varepsilon}} \, \frac{s}{1 + \tau _{\varepsilon} \, s} \, .
\ee
If we then rescale $\opO$ as follows
\be
\opL = - \frac{1 + \tau _{\varepsilon} \, s}{2 \, \sqrt{\rho \, J_{1} \, \tau _{\varepsilon}} \, s^{3/2}} \, \opO \, ,
\ee
we have that both $\opO$ and $\opL$ enjoy the same homogeneous solutions\footnote{Which are the physically significant one, as discussed in Section \ref{Sec-2}.}, provided that we keep ourself away form the singularities of $\opO$ in the Laplace domain. Clearly, this rescaled operator is expressed in terms of negative powers of $s$ as follows
	\be \label{eq-L-voigt}
	\opL \overset{s \to \infty}{\sim} \opL _{0} + \frac{1}{s^{1/2}} \, \opL _{1} + \frac{1}{s} \, \opL _{2} + 
	\frac{1}{s^{3/2}} \, \opL _{3} \, ,
	\ee
where
	\be
	\opL _{0} &\!\!=\!\!& \frac{\partial}{\partial x} \, ,\\
	\opL _{1} &\!\!=\!\!& - \frac{1}{2} \sqrt{\frac{\tau _{\varepsilon}}{\rho \, J_{1}}} \frac{\partial ^{2}}{\partial x^{2}} - \frac{1}{2} \sqrt{\frac{\rho \, J_{1}}{\tau _{\varepsilon} ^{3}}} \, , \\
	\opL _{2} &\!\!=\!\!& \frac{1}{\tau _{\varepsilon}} \, \frac{\partial}{\partial x} \, , \\
	\opL _{3} &\!\!=\!\!& - \frac{1}{2} \frac{1}{\sqrt{\rho \, J_{1} \, \tau _{\varepsilon}}} \, \frac{\partial ^{2}}{\partial x^{2}} \, .
	\ee

	Furthermore, from Eq.~(\ref{eq-L-voigt}) we can also infer that $\nu _{1} = 1/2$, $\nu_{2} = 1$, $\nu _{3} = 3/2$ and that $N=3$. Thus from the theorem by Friedlander and Keller we deduce that
	\be
	\lambda _{k} = \frac{k}{2} \, , \,\,\, k \in \N \, , \quad \mbox{and} \quad j(i, k) = k - 2 \, \nu _{i} \, , \,\,\, 
	k \in \N \, , \,\,\, i = 1, 2, 3 \, .
	\ee
	Therefore, if we proceed following the general procedure of the Buchen-Mainardi Algorithm, we can then compute the coefficients $A_{k, \ell}$. In particular, for the coefficients for which $1 \leq l \leq k$ we find that
	\be
	A_{k, \ell} = \frac{1}{2} \left( A_{k-1, \ell + 1} - 2 \, A_{k - 1, \ell} + A_{k - 1, \ell - 1} + A_{k -3, \ell +1} \right) \, .
	\ee
	
	Considering an initial step input (\ie $r_{0} (s) = 1/s$), then the function $\wt{\Phi} _{k} (s, x)$ reads
	\be
	\begin{split}
	\wt{\Phi} _{k} (s, x) &= s^{- (\lambda _{k} +1)} \, \exp \left[ - x \mu _{+} (s) \right] =\\
	&= \frac{1}{s ^{\frac{k}{2} + 1}} \, \exp \left( - x\, \sqrt{\frac{\rho \, J_{1}}{\tau _{\varepsilon}}} \, s^{1/2} \right) \, ,
	\end{split}
	\ee 
	that, inverting back to the time domain, gives
	\be
	\Phi _{k} (t, x) = t^{k/2} \, F_{1/2} \left(\sqrt{\frac{x^{2} \, \rho \, J_{1}}{\tau _{\varepsilon} \, t}}, \frac{k}{2} \right) \, .
	\ee
	At this point, we are able to write the wave-front expansion for the Voigt model, namely
	\begin{equation}
	r _{V} (t, x) \sim 
	\sum _{k=0} ^\infty \,  \sum _{\ell = 0} ^k A_{k, \ell} \, \frac{x^\ell}{\ell !}  \, 
	t^{k/2} \, F_{1/2} \left(\sqrt{\frac{x^{2} \, \rho \, J_{1}}{\tau _{\varepsilon} \, t}}, \frac{k}{2} \right)
	 \, , \quad \mbox{as} \,\, t \to 0 ^{+} \, ,
	\end{equation}
	where  the coefficients $A_{k, \ell}$ are determined as follows
\begin{equation}
 \begin{cases}
  A_{k, 0} = \delta_{k 0} \qquad \ell=0 \\
 A_{k, \ell} = \frac{1}{2} \left( A_{k-1, \ell + 1} - 2 \, A_{k - 1, \ell} + A_{k - 1, \ell - 1} + A_{k -3, \ell +1} \right) \qquad 1 \leq \ell \leq k \\
  A_{k, \ell} = 0 \qquad \ell>k \, .
 \end{cases}
\end{equation}

%%%%
%%%% Long Time VOIGT
\subsection{Long time asymptotic expansion}
	From Eq.~(\ref{eq-mu-voigt}) we can immediatly deduce that
\be
\wt{r} (s, x) = \frac{1}{s} \, \exp \left[ - x \, \mu (s) \right] = 
\frac{1}{s} \, \exp \left[ - \sqrt{J_{1} \, \rho} \, \frac{x \, s}{(1 + \tau _{\varepsilon} \, s) ^{1/2}} \right] \, .
\ee		
The inversion of this function to the time domain, as $s \to 0$ or equivalently as $t \to \infty$, can be performed by means of the saddle-point approximation method. Indeed, as discussed in \cite{Jeff-1931}, for the Voigt model we get the following long time asymptotic expansion,
\be
r (t, x) \overset{t \to \infty}{\sim} \frac{1}{2} \, \left\{ 1 + \texttt{erf} \left( \frac{t - \sqrt{J_{1} \, \rho} \,x}{\sqrt{2 \, \tau _{\varepsilon} \, t}} \right) \right\} \, .
\ee    

%%%%
\subsection{Numerical Results}
Here are shown the plots of both the asymptotic expansions for the ordinary Voigt model, computed at different fixed values of $x$. Then, we underline a numerical matching between the wave-front expansion and the long time expansion. In particular, in order to roughly show the matching, we perform a polynomial interpolation between the two approximations.

\begin{figure}[h!]
\centering
\includegraphics[width=13cm]{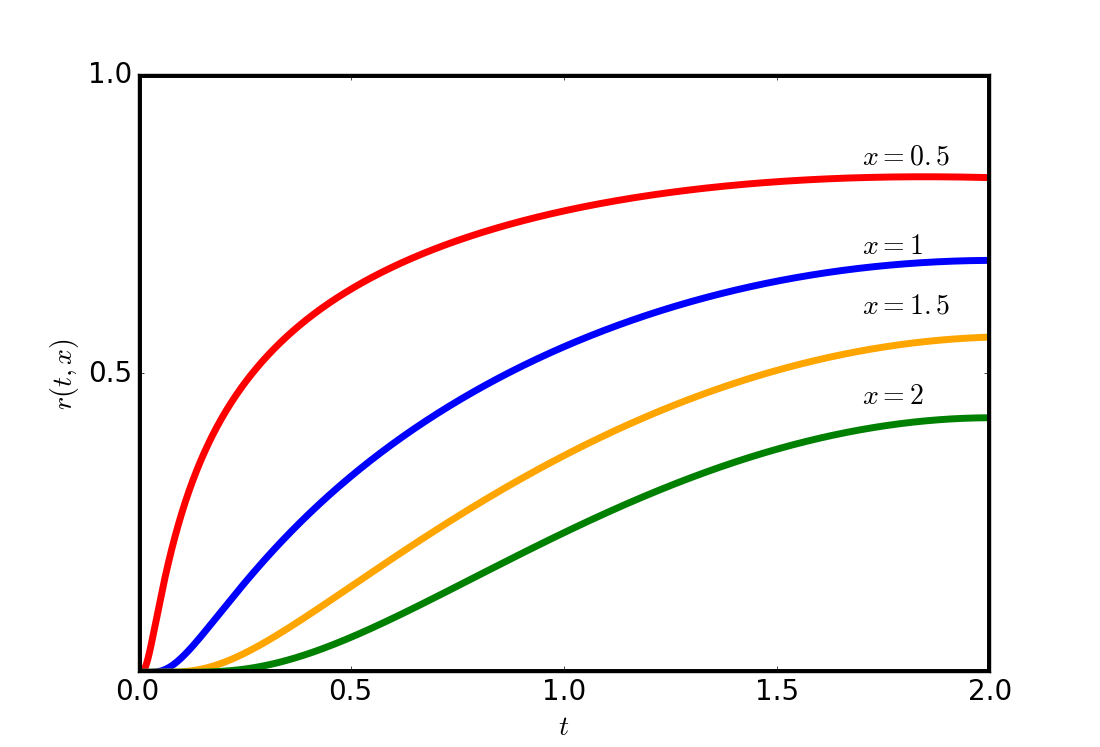}
\caption{The wave-front expansion for the ordinary Voigt model. This approximation can be trusted until values of $t\leq 2$, for which the expansion is monotonic, while we can note that this expansion is not acceptable for long times.}
\end{figure}	

\begin{figure}[h!]
\centering
\includegraphics[width=13cm]{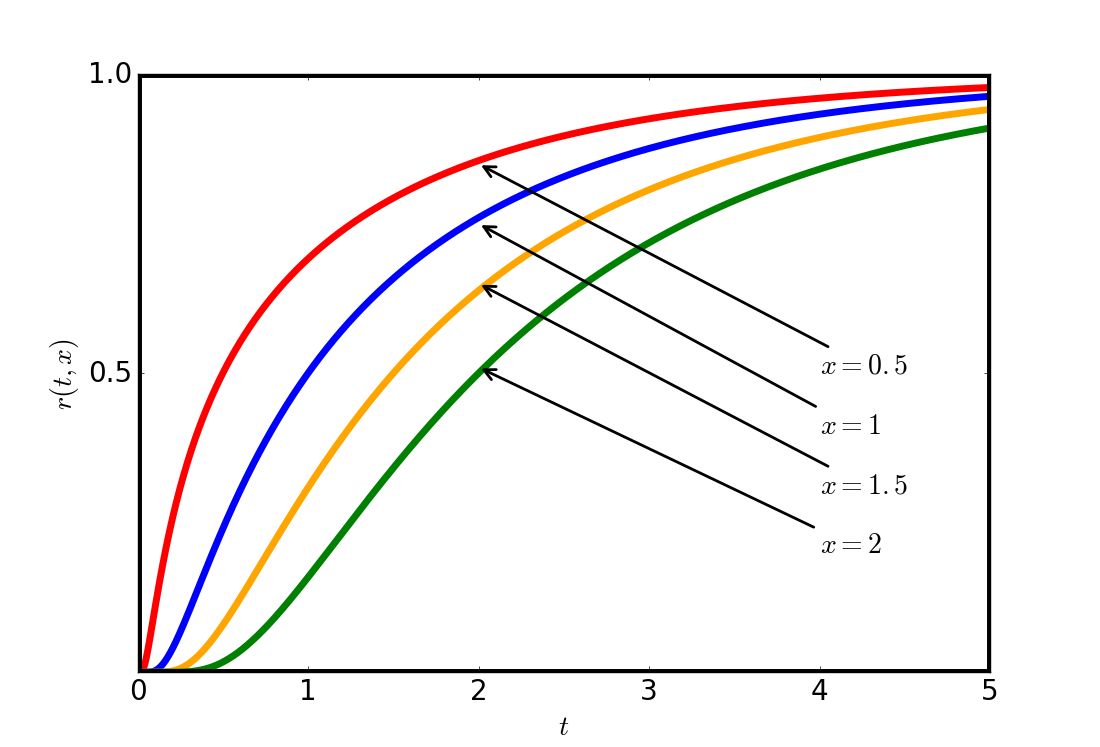}
\caption{The long time asymptotic expansion for the ordinary Voigt model.}
\end{figure}		

\begin{figure}[h!]
\centering
\includegraphics[width=13cm]{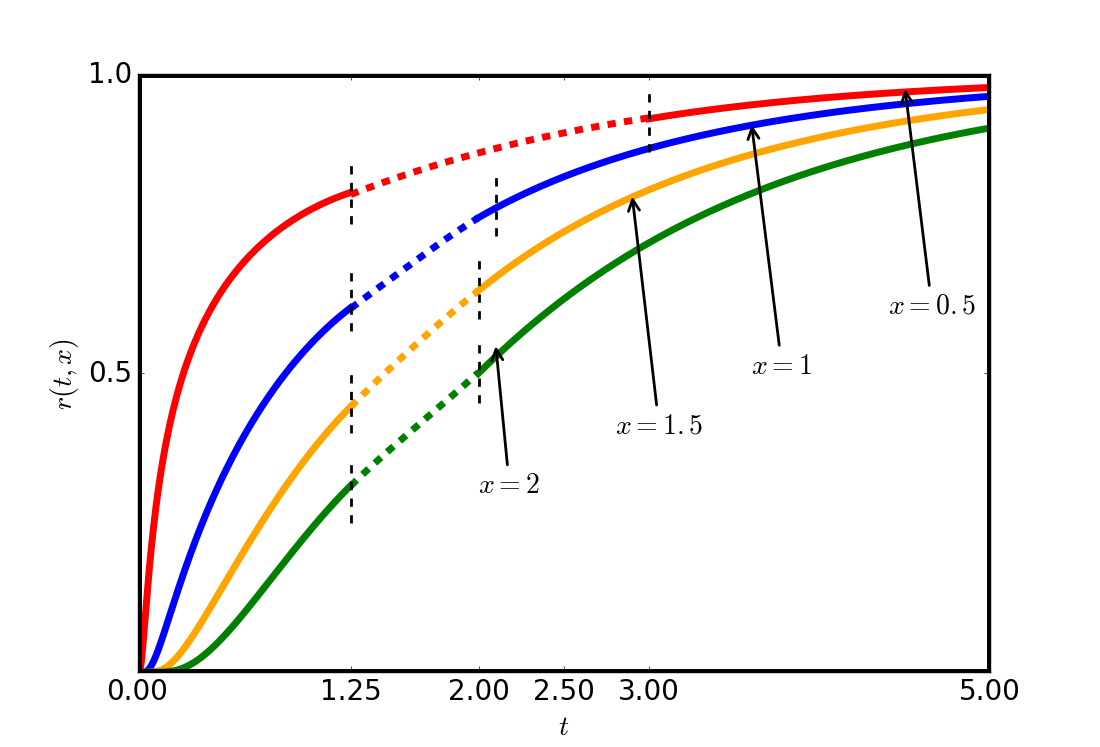}
\caption{The matching between the two asymptotic expansions for the ordinary Voigt model. The dashed line showed in the suitable intervals represents the polynomial curve that fit the two different behaviors.}
\end{figure}

\newpage

\section{The Fractional Voigt Model of order $\alpha$} \label{Sec-6}

	The constitutive equation fractional Voigt model reads
	\be
	\sigma (t) = m \, \varepsilon (t) + b_{1} \, D^\alpha _t \varepsilon (t) \, ,
	\ee
where $m \geq 0$ and $b_{1}$ is a strictly positive constant, as for the ordinary case.
	As for the fractional Maxwell model, we have replaced the ordinary time derivatives with the Caputo's fractional derivative, as defined in \eqref{eq-frac-der}.
	
	In the Laplace domain we can rewrite the constitutive equation as follows
	\be
	\wt{\sigma} (s) = ( m + b_{1} \, s^\alpha) \, \wt{\varepsilon} (s) \, ,
	\ee
from which we immediately read off the Laplace transform of the creep compliance,
	\be \label{eq-J-frac-voigt}
	\wt{J} _{V, \alpha} (s) = \frac{J_{1}}{s \, (1 + \tau _{\varepsilon} \, s^\alpha)} \, ,
 	\ee
where $J_{1} = 1/m$ and $\tau _{\varepsilon} = b_{1} / m$.

\subsection{Wave-front expansion for the general case}

Following the same procedure extensively discussed above, we have
\be
\mu (s) = \sqrt{\rho} \, s	\, \left[s \, \wt{J} _{V, \alpha} (s) \right] ^{1/2} =
\sqrt{\frac{\rho}{b_{1}}} \,s^{1-\alpha/2}\, \left[
1+	\frac{m}{(s\, b_1)^\alpha}	\right]^{-1/2} \,,
\ee	

Now, as $s \to \infty$ we find that
	\be
	\mu (s) = \sqrt{\frac{\rho}{b_{1}}}
	\sum _{n=0}^\infty  \binom {-1/2} {n}
	\left(\frac{m}{b_1}\right)^n
	s^{1-\alpha(n+1/2)} \,.
	\ee	
Again, at this point, we distinguish different cases as shown for the fractional Maxwell model. As $1-\alpha(n+1/2) \geq 0$, then we have to satisfy the condition $n\leq [1/\alpha - 1/2]$.

In this paper, we only consider the fractional Voigt model of order $\alpha = 1/2$, so that $n\leq 2$ in our expansion. 

\subsection{The Fractional Voigt Model of order 1/2}
As said before, fixing $\alpha=1/2$,
Eq.~\eqref{eq-J-frac-voigt} turns into
	\be \label{eq-J-voigt_1/2}
	\wt{J} _{V, 1/2} (s) = \frac{J_{1}}{s \, (1 + \tau _{\varepsilon} \, s^{1/2})} \, .
 	\ee

\subsubsection{Wave-front Expansion}
Then, 
\be \label{eq-mu-voigt1/2}
\mu (s) = \sqrt{\rho} \, s	\, \left[s \, \wt{J} _{V, 1/2} (s) \right] ^{1/2} =
\sqrt{\frac{\rho}{b_{1}}} \,s^{1-\alpha/2}\, \left[
1+	\frac{m}{(s\, b_1)^{1/2}}	\right]^{-1/2} \,,
\ee	
that, as $s \to \infty$, reads
	\be \label{eq-exp-mu-voigt-1/2}
	\mu (s) = \sqrt{\frac{\rho}{b_{1}}}
	\left[
	s^{3/4} - \frac{m}{2\, b_1} s^{1/4} +
	\sum _{n=2}^\infty  \binom {-1/2} {n}
	\left(\frac{m}{b_1}\right)^n
	s^{(3-2n)/4}
	\right] \,.
	\ee	

	From \eqref{eq-exp-mu-voigt-1/2} it is straightforward to evaluate the expression for $\mu _+ (s)$, namely
	\be
	\mu_+(s) = \sqrt{\frac{\rho}{b_{1}}}
\left(
s^{3/4} - \frac{m}{2\, b_1} s^{1/4}
\right) \,,
	\ee
and, consequently,
	\be
	\mu ^{2} (s) &\!\!=\!\!& \frac{\rho}{b_{1}} \, \frac{s^{3/2}}{1 + \frac{m}{b_1 \, s^{1/2}}} = \rho \frac{s^2}{m+b_1 \, s^{1/2}} \, , \\
	\mu ^{2} _{+} (s) &\!\!=\!\!& \frac{\rho}{b_1} \, \left( s^{3/2} + \frac{m^2}{4\, b_1^2}s^{1/2} - \frac{m}{b_1}s \right) \, ,
	\ee
thus, we get
	\be
	\mu ^{2} (s) - \mu _{+} ^{2} (s) = 
	\frac{\rho}{4\, b_1^3 \, ({m+b_1 \, s^{1/2}})} \left[ 3b_1 m^2 s - m^3 s^{1/2} \right] \,.
	\ee
	
	Therefore, the differential operator $\opO$ is given by
\be
\opO &\!\!=\!\!& \frac{\partial ^{2}}{\partial x ^{2}} - 2 \,  \sqrt{\frac{\rho}{b_{1}}}
\left(s^{3/4} - \frac{m}{2\, b_1} s^{1/4} \right) \frac{\partial}{\partial x}
-\frac{\rho \, \left( 3b_1 m^2 s - m^3 s^{1/2} \right)}{4\, b_1^3 \, ({m+b_1 \, s^{1/2}})} \, .
\ee
Then, rescaling $\opO$, we find $\opL$, \ie
\be \label{eq-L-frac-voigt}
	\opL \overset{s \to \infty}{\sim} \opL _{0} + \frac{1}{s^{1/2}} \, \opL _{1} + \frac{1}{s^{3/4}} \, \opL _{2} + 
	\frac{1}{s} \, \opL _{3}+
	\frac{1}{s^{5/4}} \, \opL _{4} \, ,
	\ee
where
	\be
	\opL _{0} &\!\!=\!\!& \frac{\partial}{\partial x} \, ,\\
	\opL _{1} &\!\!=\!\!& - \frac{1}{2\, \tau^{1/2}}\frac{\d}{\d x} \, , \\
	\opL _{2} &\!\!=\!\!& -\sqrt{\frac{\mu \tau^{1/2}}{2\rho}} \frac{\d ^2}{\d x^2} - \sqrt{\frac{2\rho}{\mu}}\frac{1}{16\tau} \, , \\
	\opL _{3} &\!\!=\!\!& -\frac{1}{2\tau} \frac{\d}{\d x}  \, ,\\
	\opL _{4} &\!\!=\!\!&  -\sqrt{\frac{\mu}{2\rho\tau ^{1/2}}}\frac{\d ^2}{\d x ^2} + \frac{2\rho}{\mu} \frac{3}{16\, \tau ^{5/4}} \, .
	\ee

%%%%%%%%%%%%%%CP

Moreover, from Eq.~(\ref{eq-L-frac-voigt}) we can also infer that $\nu _{1} = 1/2$, $\nu_{2} = 3/4$, $\nu _{3} = 1$, $\nu_4 = 5/4$ and that $N=4$. Again, thanks to the theorem by Friedlander and Keller we deduce that
	\be
	\lambda _{k} = \frac{k}{4} \, , \,\,\, k \in \N \, , \quad \mbox{and} \quad j(i, k) = k - 4 \, \nu _{i} \, , \,\,\, 
	k \in \N \, , \,\,\, i = 1, 2, 3,4 \, .
	\ee
	Hence, the coefficients $A_{k, \ell}$ are easily computed. Specifically, the coefficients for which $1 \leq l \leq k$ are given by
	\be
	A_{k, \ell} = - \Bigg( 
	\frac{1}{2\tau ^{1/2}} A_{k-2, \ell} - \sqrt{\frac{\mu \tau ^{1/2}}{2\rho}} \, A_{k - 3, \ell +1} 
	-\sqrt{\frac{2\rho}{\mu}}\frac{1}{16\tau} A_{k - 3, \ell - 1} +\\ 
	-\frac{1}{2\tau} A_{k -4, \ell}
	-\sqrt{\frac{\mu}{2\rho\tau ^{1/2}}} A_{k -5, \ell +1} + \sqrt{\frac{2\rho}{\tau}} \frac{3}{16\tau ^{5/4}} A_{k -5, \ell -1} \Bigg) \, . \notag
	\ee
	
	Considering an initial step input (\ie $r_{0} (s) = 1/s$), then the function $\wt{\Phi} _{k} (s, x)$ is given by
	\be
	\begin{split}
	\wt{\Phi} _{k} (s, x) &= s^{- (\lambda _{k} +1)} \, \exp \left[ - x \mu _{+} (s) \right] =\\
	&= s^{-1-k/4} \, \exp \left[ - x\, \sqrt{\frac{\rho}{b_1}} \,\left( s^{3/4} - \frac{m}{2\, b_1} s^{1/4} \right) \right] \, ,
	\end{split}
	\ee 
	that it is not easy to invert back, at least analytically, to the time domain.

\subsubsection{Long time asymptotic expansion}
	From Eq.~(\ref{eq-mu-voigt1/2}) one can easily infer that
\be\label{LT-fv-pre}
\wt{r} (s, x) = \frac{1}{s} \, \exp \left[ - x \, \mu (s) \right] = 
\frac{1}{s} \, \exp \left[ - x \sqrt{\frac{\rho}{2\mu}}\, \left(s -\frac{b_1}{4\mu} s^{3/2}\right) \right] \, .
\ee		
The inversion of this function to the time domain, as $s \to 0$ or equivalently as $t \to \infty$, appears to be particularly involved. Nevertheless, if we consider only the first term of the expansion, we get the following long time asymptotic behaviour,
\be\label{LT-fv}
r (t, x) \overset{t \to \infty}{\sim} \frac{1}{2} \, \Theta \left( t-x\sqrt{\frac{\rho}{2\mu}} \right) \, ,
\ee    
where $\Theta$ is the Heaviside step function. A numerical inversion of Eq.~\eqref{LT-fv-pre} has shown negligible deviations from Eq.~\eqref{LT-fv} once further terms of the expansion are taken into account.

\section{Conclusions}
	In this article we revisited and generalized the wave-front formalism for viscoelatic transient phenomena, first proposed by Buchen and Mainardi in \cite{Buchen-Mainardi 1975}, also providing a wide variety of applications to the most relevant linear viscoelastic models.
	
	Specifically, in Section \ref{Sec-2} we outlined a way to set the problem of the production and propagation of transient waves in (linear) viscoelastic media. Then, we precisely derived the Buchen-Mainardi algorithm in a very general fashion, also laying out the procedure for generic (linear) media.  
	In Section \ref{Sec-3} we then performed the wave-front expansion for an ordinary Maxwell solid considering, as source of the perturbations of the material, an instantaneous step input. Subsequently, we computed the long time response of the material and we compared the plots of the two approximations. The result of this analysis is a nice matching of the two asymptotic expansions. 
	
	In Section \ref{Sec-4} we then tackled the same problem in the framework of the fractional Maxwell model. The section begins with a discussion of the wave-front expansion for a generic fractional Maxwell model of order $\alpha \in \R ^+$. Then, in order to be able to perform some explicit computations, we decided to inspect two realizations of the model, respectively $\alpha = 1/2$ and $\alpha = 3/4$. Although this analysis shows that we still have a quite nice matching between the short time and the long time approximations, this seems to emerge on a time scale which is way shorter than the one that we found for the ordinary model. Nevertheless, the matching still appears within the range of validity of the wave-front approximation (which is obtained from a truncation of the series in Eq.~\eqref{eq-fin} to the 30th term).
	
	In Section \ref{Sec-5} we turned our attention to the ordinary Voigt model. As above, after computing both the wave-front expansion and the long time approximation (the latter was obtained by Jeffreys in \cite{Jeff-1931}), we graphically compared the two asymptotic behaviors. Unfortunately, the wave-front expansion breaks way before showing any explicit matching, and this is independent from the number of terms considered for the series in Eq.~\eqref{eq-fin} (aside from minor corrections). 
	
	Finally, in Section \ref{Sec-6} we performed the same procedure for the fractional Voigt model of order $1/2$. We decided not to show plots for the two approximations because, as already seen in the Maxwell models, the wave-front expansion breaks even earlier than the one of the ordinary model. Moreover, the long time approximation shows very little deviations from the step function.
	
	As a final remark, it is important to stress that the long time expansions presented in this paper have been obtained following a very simplified approach. The reason for this is that the main focus of the paper is on the (short time) wave-front expansion, and the comparison between the two asymptotic approximation is used to qualitatively infer to which extent the latter can be trusted, in some very simple cases. Clearly, if we wanted to provide a more precise description of the long time expansion we should recur to some more sophisticated such as the one described in \cite{Stronzo2, Pipkin}. Furthermore, it is also important to remark, for sake of completeness, that in \cite{Stronzo2, Pipkin} the authors address an important result by Kolsky \cite{K} concerning a peculiar universality of viscoelastic pulse shapes.

	\section*{Acknowledgments}
	The authors acknowledge the two anonymous reviewers for the constructive comments and suggestions which have helped to improve the manuscript significantly.
	
	The work of the authors has been carried out in the framework of the activities of the National Group of Mathematical Physics (GNFM, INdAM).

	Moreover, the work of A.G. has been partially supported by \textit{GNFM/INdAM Young Researchers Project} 2017 ``Analysis of Complex Biological Systems''.

%%%%End of the main text
%%%%%%%%%%%%%%%
	
 %%%%%% Bibliography

 \end{document}